\documentclass[12pt]{article}
\pdfoutput=1
\usepackage{amssymb}
\usepackage{amsmath}
\usepackage{amsfonts}
\usepackage{graphicx}
\usepackage{mathrsfs}
\usepackage{comment}
\usepackage{cite}
\usepackage{hyperref}
\usepackage[normalem]{ulem}
\usepackage{color}
\begin{document}

\newcommand{\BB}[1]{\textbf{\textcolor{blue}{[BB: #1]}}}
\newcommand{\WY}[1]{\textbf{\textcolor{red}{[WY: #1]}}}
\newcommand{\wy}[1]{\textbf{\textcolor{red}{#1}}}

\renewcommand{\figurename}{Fig.}
\renewcommand{\tablename}{Table.}
\newcommand{\Slash}[1]{{\ooalign{\hfil#1\hfil\crcr\raise.167ex\hbox{/}}}}
\newcommand{\bra}[1]{ \langle {#1} | }
\newcommand{\ket}[1]{ | {#1} \rangle }
\newcommand{\beq}{\begin{equation}}  \newcommand{\eeq}{\end{equation}}
\newcommand{\bef}{\begin{figure}}  \newcommand{\eef}{\end{figure}}
\newcommand{\bec}{\begin{center}}  \newcommand{\eec}{\end{center}}
\newcommand{\non}{\nonumber}  \newcommand{\eqn}[1]{\begin{equation} {#1}\end{equation}}
\newcommand{\laq}[1]{\label{eq:#1}}  
\newcommand{\dd}[1]{{d \o d{#1}}}
\newcommand{\Eq}[1]{Eq.(\ref{eq:#1})}
\newcommand{\Eqs}[1]{Eqs.(\ref{eq:#1})}
\newcommand{\eq}[1]{(\ref{eq:#1})}
\newcommand{\Sec}[1]{Sec.~\ref{chap:#1}}
\newcommand{\ab}[1]{\left|{#1}\right|}
\newcommand{\vev}[1]{ \left\langle {#1} \right\rangle }
\newcommand{\bs}[1]{ {\boldsymbol {#1}} }
\newcommand{\lac}[1]{\label{chap:#1}}
\newcommand{\SU}[1]{{\rm SU{#1} } }
\newcommand{\SO}[1]{{\rm SO{#1}} }
\def\({\left(}
\def\){\right)}
\def\dt{{d \o dt}}
\def\diag{\mathop{\rm diag}\nolimits}
\def\Spin{\mathop{\rm Spin}}
\def\O{\mathcal{O}}
\def\U{\mathop{\rm U}}
\def\Sp{\mathop{\rm Sp}}
\def\SL{\mathop{\rm SL}}
\def\tr{\mathop{\rm tr}}
\def\ebq{\end{equation} \begin{equation}}
\newcommand{\OR}{~{\rm or}~}
\newcommand{\AND}{~{\rm and}~}
\newcommand{\EV}{ {\rm \, eV} }
\newcommand{\KEV}{ {\rm \, keV} }
\newcommand{\MEV}{ {\rm \, MeV} }
\newcommand{\GEV}{ {\rm \, GeV} }
\newcommand{\TEV}{ {\rm \, TeV} }
\def\o{\over}
\def\a{\alpha}
\def\b{\beta}
\def\c{\varepsilon}
\def\d{\delta}
\def\e{\epsilon}
\def\f{\phi}
\def\g{\gamma}
\def\h{\theta}
\def\k{\kappa}
\def\l{\lambda}
\def\m{\mu}
\def\n{\nu}
\def\p{\psi}
\def\q{\partial}
\def\r{\rho}
\def\s{\sigma}
\def\t{\tau}
\def\u{\upsilon}
\def\v{\varphi}
\def\w{\omega}
\def\x{\xi}
\def\y{\eta}
\def\z{\zeta}
\def\D{\Delta}
\def\G{\Gamma}
\def\H{\Theta}
\def\L{\Lambda}
\def\F{\Phi}
\def\P{\Psi}
\def\S{\Sigma}
\def\me{\mathrm e}
\def\ol{\overline}
\def\tl{\tilde}
\def\*{\dagger}

\newcommand{\changed}[2]{{\protect\color{red}\sout{#1}}{\protect\color{blue}\uwave{#2}}}

\thispagestyle{empty}
\begin{center}
\hfill TU-1168, PITT-PACC-2403\\

\vspace{1.3cm}

{\Large\bf 
Cosmic Stability of Dark Matter \\ 
from Pauli Blocking
}
\vspace{1.3cm}

{\bf Brian Batell$^{1}$ and Wen Yin$^{2,3,4,5}$}

\vspace{10pt}
\vspace{0.5cm}
{\em 
$^{1}$ Pittsburgh Particle Physics, Astrophysics, and Cosmology Center, Department of Physics and Astronomy, University of Pittsburgh, Pittsburgh, USA \\
$^{2}$ Department of Physics, Tohoku University, 
Miyagi 980-8578, Japan\\
$^{3}$ Department of Physics,  Tokyo Metropolitan University,  
Tokyo, 192-0397, Japan\\
$^{4}$ Department of Physics, Faculty of Science, The University of Tokyo,  \\ 
Bunkyo-ku, Tokyo 113-0033, Japan\\
 $^{5}${ Department of Physics, KAIST, Daejeon, Korea} \vspace{5pt}
\\}
\vspace{0pt}
\vspace{0.5cm}
\date{\today $\vphantom{\bigg|_{\bigg|}^|}$}
\abstract{ 
Why does dark matter (DM) live 
longer than the age of the Universe? 
Here we study a novel 
sub-eV
scalar DM candidate whose stability is due to
the Pauli exclusion of its fermionic decay products. 
We 
analyze the stability of the DM condensate against 
decays, scatterings (i.e., evaporation), and parametric resonance, 
delineating 
the viable parameter regions in which DM is cosmologically stable.  
In a minimal scenario in which 
the scalar DM decays to   
 a pair of new exotic fermions, 
we find that 
scattering can 
populate an interacting thermal dark sector component
to 
energies far
above the DM mass. 
This self-interacting dark radiation may potentially 
alleviate the Hubble tensions.  Furthermore, our scenario can  
be probed 
through precise measurements of the halo mass function or the
 masses of dwarf spheroidal galaxies since  scattering prevents the DM from becoming too dense. 
 On the other hand, if the lightest neutrino 
stabilizes the DM, 
the cosmic neutrino background (C$\nu$B) can be significantly altered from the $\L$CDM prediction and thus be probed in the future by C$\nu$B detection experiments.}
\end{center}
\clearpage

\setcounter{page}{1}
\setcounter{footnote}{0}

\setcounter{footnote}{0}
\section{Introduction}
The stability of dark matter (DM) on cosmic timescales provides an important clue to its fundamental nature. 
For example, it may suggest DM is endowed with a conserved charge, has a small mass, and/or has suppressed interactions, leading to a small or vanishing decay rate. 
Understanding the possible reasons behind DM stability is extremely important because they may suggest novel correlated observational signatures, as is well illustrated by the cases of DM stabilized by symmetry (e.g., WIMP DM) or by small mass/suppressed interactions (e.g., ALP DM).

In this paper, we explore a novel explanation for the stability of DM in our Universe that exploits the Pauli exclusion principle. In a simple model of 
sub-eV scalar DM with sizable Yukawa couplings to even lighter or massless exotic fermions, the DM proper lifetime in vacuum is much shorter than the age of the Universe. However, the fermionic states accessible in the decay are rapidly filled,
thereby Pauli blocking the decay and effectively rendering DM stable on cosmological timescales. 
We carefully study the stability of the scalar DM condensate by taking account of decay, scattering, and parametric resonance effects, which turn out to be important during different eras of the DM density evolution. 
In particular, we highlight
the importance of DM scattering, including the self-annihilation of condensate particles and their scattering with thermalized scalars, which proceeds via the same coupling that mediates 
the DM decay. Such scattering processes can lead to the fast evaporation of the condensate and thus significantly constrain this scenario at large couplings. Nevertheless, we demonstrate that over a broad range of DM masses and at moderate values of the coupling constant, a viable scenario of DM stabilization due to Pauli blocking 
can be realized.  

For relatively sizable Yukawa couplings, our scenario  leads to a novel late time cosmology. 
The decay and scattering of DM can lead to the formation of 
a quasi-thermal component of fermions and scalars that behaves as dark radiation with a temperature far above the mass of the DM. 
The resulting dark radiation can have a substantial abundance and, furthermore, is 
 self-interacting 
since it attains kinetic equilibrium via the decay and inverse decay of the relativistic dark sector plasma. 
The deviation of the effective neutrino number at recombination can be as large as $\O(0.1)$ with only small corrections to the DM abundance relative to $\L$CDM. 
Very interestingly, such a scenario may have important implications for the 
more than $5\sigma$ tension between local and global measurements of the Hubble constant~\cite{Planck:2018vyg,Riess:2011yx, Bonvin:2016crt,Riess:2016jrr, Riess:2018byc,Birrer:2018vtm,Riess:2021jrx}. (See also Refs.~\cite{Freedman:2020dne,Khetan:2020hmh}, and reviews~\cite{Freedman:2017yms, Verde:2019ivm,DiValentino:2020zio,Schoneberg:2021qvd} for the Hubble tension.) 
Previous studies have shown that 
the 
Hubble tensions may be alleviated due to the presence of self-interacting dark radiation~\cite{Mortsell:2018mfj,Riess:2019cxk,DiValentino:2020zio,Blinov:2020hmc}. 
Furthermore, even if the dark sector only couples to the SM via gravity, the resulting DM halo density profiles will feature a cutoff since the formation of dense regions of DM will be obstructed by fast DM evaporation.
{Thus our scenario may prefer
 cored DM density profiles with a cutoff depending on the masses and coupling strength.
Precisely simulating structure formation and measuring the small-scale structure, e.g., via halo-mass function measurements in Vera Rubin Observatory~\cite{LSSTDarkMatterGroup:2019mwo} or 21 cm line observation~\cite{Sitwell:2013fpa}, can thus provide probes of this scenario. 

An obvious candidate for the fermions in our scenario are the Standard Model (SM) neutrinos, which leads to several additional consequences in comparison to the case of an exotic BSM fermion.
In this case, the composition of the cosmic neutrino background (C$\nu$B) may be altered if the coupling of the DM to neutrinos is not diagonal in the neutrino mass eigenbasis. 
This is because the heavier neutrinos may decay into DM and a lighter neutrino. The produced DM, which is energetic and thus 
not Pauli blocked, can further decay into lighter neutrinos. The cascade decays result in a C$\nu$B that is composed entirely of the lightest neutrinos, some of which are boosted. 
Interestingly, in the case of  normal {ordering}
our scenario predicts up to a 
factor 1-10 enhancement of the 
boosted neutrino
interaction rate to an electron by the neutrino decays with an energy of $\O(0.01)\EV$.
{In addition, either in the normal or inverted ordering, the DM may further produce the lightest neutrinos via the evaporation or decay while still maintaining stability of the DM condensate.
For certain regions of parameter space, although the typical energy of each neutrino is smaller than $\O(0.01)\EV$,} this contribution can enhance the C$\nu$B interaction rate by many orders of magnitude.
Thus, the imprint of DM interactions on the C$\nu$B may potentially be probed in future experiments such as PTOLEMY~\cite{Betti:2019ouf, McKeen:2018xyz, Long:2014zva}.

Before moving to the main part of the paper, let us mention a few relevant studies.
Ref.~\cite{Bjaelde:2010vt} investigated the possibility of scalar DM decaying to neutrinos and the impact of Pauli blocking. However, this work did not consider the effect of scattering, which turns out to be important even for modest Yukawa couplings. 
 The cosmological production of fermions due to 
 scalar field dynamics,
 either from broad parametric resonance~\cite{Greene:1998nh,Greene:2000ew} or perturbative decays~\cite{Randall:2015xza,Moroi:2020has,Moroi:2020bkq},
 has been widely investigated in the context of inflationary reheating or preheating. 
While distinct from our scenario, light dark sectors in the form of a 
nearly degenerate Fermi sea
has been explored in Refs.~\cite{Randall:2016bqw,Moroi:2020has,Davoudiasl:2020uig,Carena:2021bqm}. 

This paper is organized as follows. In the next section, we discuss our theoretical setup and in the perturbative regime.
We formulate a system of Boltzmann equations, including quantum statistical effects, which allows us to account for the Pauli blocking effect and investigate the stability of the DM condensate. (Some technical details are presented in Appendices~\ref{app:1} and \ref{app:2}).
In \Sec{DMODS}, we investigate the minimal scenario in which DM is stabilized by exotic BSM fermions.
We study the cases of DM annihilation with and without a chirality suppression, the {production of} dark radiation and the DM density evolution, the potential implications for the Hubble tensions, the cutoff of the present local DM density, the broad parametric resonance regime, and conclude with a summary of the cosmology of this scenario. 
In \Sec{DMONS} we explore the scenario in which the C$\nu$B stabilizes  DM and the resulting implications for the evolution and detection of the 
the C$\nu$B. 
(A gauge-invariant UV model is discussed in Appendix~\ref{sec:ap1}) 
{In \Sec{gene} we discuss some generalizations to other models and signatures.}
The final section is devoted to the conclusions and outlook.

\section{Setup and Boltzmann equations}

\subsection{Setup}
\lac{DMdecay}
To clarify our idea, we introduce a minimal model containing a real scalar field $\f$ with mass $m_\f$, which will be identified as the DM, and a 
Majorana fermion $\psi$ with mass $m_\psi$.
Throughout this work we will consider 
{$m_\psi 
<
m_\f/2$}, and for the moment we  neglect the effect of the fermion mass.\footnote{We note that fermionic bound states~\cite{Afshordi:2005ym,Bjaelde:2010vt,Gogoi:2020qif,Smirnov:2022sfo} do not form in our scenario since the range of the Yukawa force is much smaller than the fermion de Broglie wavelength. 
}
Here and in \Sec{DMODS}
 we assume that $\p$ is a BSM fermion with negligible couplings
to SM particles, while in \Sec{DMONS} we will examine the case in which $\psi$ are the SM neutrinos. 
The scalar-fermion interaction is represented by the Lagrangian 
\beq
\laq{int}
{\cal L}\supset - \frac{y}{2}\f   \bar{\psi}  \psi \, ,
\eeq
where $y$ is a real Yukawa coupling.\footnote{ 
We have also 
considered the case of a pseudoscalar coupling, see Appendix~\ref{app:1}. 
The dynamics do not change much, but the annihilation of $\f$ condensates proceeds in the $D$-wave, and the dark radiation production is similar to the chirality-suppressed case examined in \Sec{cs} even if $m_\psi=0$. 
}
As a consequence of this interaction, $\f$ decays to a $\psi$ pair in the vacuum. 

Suppose that at the redshift $z=z_{\rm ini}$,  there is a population of non-relativistic $\f$ particles with number density $n_\f \neq 0$ and negligible population of $\p$ particles, $n_\p=0$. 
Then, $\f$ decays via the process  
$\f \to  \p\p$,
with decay rate  
\begin{equation}
\label{eq:phi-decay-width-1}
 \Gamma_{\phi\to \psi\psi} \simeq 
\frac{y^2 m_\phi}{16 \pi}. 
\end{equation}
For simplicity 
we neglect the mass of $\psi$ here and in the following (see Appendix~\ref{app:1} for the formula without this approximation).
If the decay is fast enough, the states of $\p$ in the spherical shell of momentum around $E_\psi\simeq m_\f/2$ are soon occupied. 
Strictly speaking, there are also effects due to scattering processes, e.g., $\f\f\to  \psi\psi$, which will turn out to be important and will be the main focus of this paper. 
The width of the shell, $\D E$, depends on the velocity dispersion of $\f$, for which we may safely assume $\D E \ll m_\f/2$ if $\f$ is cold enough. For concreteness, we assume  that $\f$ is produced through the misalignment mechanism and that $\D E$ is negligible~\cite{Preskill:1982cy,Abbott:1982af,Dine:1982ah}.\footnote{Strictly speaking, the momentum distribution has a width of $Q m_\f$ in the misalignment case~\cite{Greene:1998nh,Greene:2000ew}, where $Q$ is defined in \Eq{con1}. } 
Our setup will work with other DM production mechanisms as well. 

Then the backreaction, i.e., Pauli blocking and inverse decay, becomes important. 
By accounting for the fermion phase space distribution, $f_\p[E_\p]$, the effective decay rate of the DM is corrected as (see also Eq.~(\ref{eq:C-phi-decay-C}) below)
\beq
\laq{decay}
\G^{\rm eff}_{\f\to \p\p}\simeq \frac{y^2 m_\phi }{16\pi} (1-2f_\p[m_\f/2]).
\eeq
Consequently $f_\p$ cannot increase beyond $f_\p=1/2$ around which the $\f$ decay is forbidden, i.e., the decay is Pauli blocked~\cite{Bjaelde:2010vt}. 

Since the Universe expands, 
the energy of the Fermi shell decreases as $
E_\psi\propto a^{-1}.
$
As the initial shell shrinks, the $\phi$ decay continually replenishes the shell of $E_\psi\simeq m_\f/2$. 
As a consequence, a distribution of $f_\p$ at $z< z_{\rm ini}$ is formed as 
\beq
\laq{step}
f_\p = \frac{1}{2}\Theta\left[\frac{m_\f}{2}-p\right]\times \Theta\left[p- \frac{m_\f}{2}\frac{ (1+z)}{(1+z_{\rm ini})}\right]\to \frac{1}{2}\Theta\left[\frac{m_\f}{2}-p\right],
\eeq
 where $\H$ is a step function.  For a sufficiently long period, $z\ll z_{\rm ini}$, the $\p$ form a Fermi sea, characterized by a Fermi sphere of energy $E_F\simeq m_\f/2$ as shown in the r.h.s.  of Eq.~\eq{step}.

Even though the DM is continuously decaying to fill the Fermi sea, the energy transferred from the DM condensate to the Fermi sea is negligible if\footnote{
The motion and gravitational potential, e.g. during structure formation, may slighly alter the momenta of $\psi$. As long as the effects do not change the $\psi$ phase distribution by a substantial amount, our conclusions do not change as long as \Eq{massb} is satisfied.}
\beq
\laq{massb}
\rho_{\f}
\gg \frac{g_\p}{2}\times \frac{E_{F}^4}{8\pi^2}\simeq g_\p \frac{m_\f^4}{256\pi^2},
\eeq
where the r.h.s is the energy density of the Fermi sphere and $g_\p=2$ denotes the spin degree of freedom of $\p.$  
For $\f$ to account for the cosmologically stable DM, this relation should be at least satisfied in the present Universe. 
By taking $\r_\f=\rho_{\rm DM} \simeq (2\,{\rm meV})^4,
$ which is the DM energy density today~\cite{Aghanim:2018eyx}, we obtain the following constraint on the DM mass in our scenario:
\beq
\laq{massbound}
m_\f \ll 0.01\EV.
\eeq
Therefore, what we will mainly discuss is a sub-eV DM candidate. 
In the following, we will see that 
additional reactions beyond decays can lead to the formation of a quasi-thermal component of scalars and fermions with temperature much higher than $m_\f$, which, however, may still be subdominant to the DM.

So far we have discussed the stability in the leading order approximation of $y$. 
As we have mentioned several times, a higher-order effect, i.e.,  $2$-to-$2$ scatterings with rates scaling as $\O(y^4)$, is quite important for discussing the stability of the DM condensate.
If either the decay or scattering (including DM annihilation) were absent and the condition \Eq{massb} is parametrically satisfied, 
 there would be both comoving number and energy conservations\footnote{In the ${1\leftrightarrow 2}$ process, $n_\f +n_\p/2$ is conserved, and in the $2\leftrightarrow 2$ process $n_\f +n_\p$ is conserved with $n_\p$ being the fermion+antifermion number. } which stabilize the DM condensate. 

The annihilation/scattering rate of a plasma $\f$ or $\psi$ particle of energy $E\gg m_\f$ with a condensate $\phi$,
 neglecting the Pauli blocking effect, is given by 
\beq
\laq{condsca}
v\s_0 n_\f \approx \frac{y^4}{\pi m_\f E} n_\f .
\eeq
where $\sigma_0$ is scattering cross section.
For now, let us consider the regime 
\beq
\laq{conHub}
\G_{\f\to \p\p} \frac{m_\f}{E}\gg v\s_0 n_\f \gg H, 
\eeq 
where the l.h.s is the (inverse) decay rate of the relativistic $(\psi) \f$ of energy $E$. 
The first inequality leads to the condition
\beq
\laq{con1}
Q\equiv \frac{y^2 \rho_\f }{m_\f^4}\ll 1.
\eeq
This corresponds to the narrow resonance regime in the context of parametric resonance~\cite{Greene:1998nh,Greene:2000ew}.
We will come back to the broad resonance regime, which can be important at early times, in Sec.\ref{secpr}. 
In the narrow resonance regime, one can study the system, including the effects of both decays/inverse decays and scatterings, by employing Boltzmann equations.

Before moving on, we remark that light scalar field will generically receive radiative corrections to its potential which are sensitive to UV physics. These corrections could be small if the dark scalar and fermion have only weak couplings to UV degrees of freedom. Alternatively,  one can consider $\phi$ to be a pseudo-Goldstone boson with interaction $\partial_\mu \phi \overline \psi \gamma^\mu\gamma^5 \psi$ instead of \Eq{int}, in which case the corrections to the potential are protected by a shift symmetry. 
Additionally, the free energy density of fermions sources a contribution to the effective potential of the scalar field. While this contribution is always subdominant in comparison to the scalar mass term for the minimal model considered here and in Sec.~\ref{fig:DMODS}, it can be relevant at early times when the fermion is the SM neutrino due to the presence of the C$\nu$B, see Sec.~\ref{fig:DMONS} for further discussion.

\subsection{Boltzmann Equations}

We now investigate the impact of scattering between the dark sector particles, which can cause the DM condensate to evaporate and lead to the formation of a thermalized dark sector component that serves as dark radiation. To investigate the stability of the DM, estimate its lifetime, and discern the properties of the dark radiation component, we will use the formalism of Boltzmann equations. As will be argued in \Sec{DMODS}, in large regions of parameter space it is expected that the system reaches a quasi-equilibrium state, which motivates us to employ a specific ansatz for the form of the phase space distribution functions that separates the condensate and the thermalized components. 
For the latter, we take for the dark radiation $\psi$ and $\phi$ particles the standard equilibrium Fermi-Dirac and Bose-Einstein distributions, which should provide a qualitatively correct description as the system thermalizes. 

We wish to follow the evolution of the phase space distributions of the scalar and fermion fields, $f_\phi[p_\phi, t]$ and $f_\psi[p_\psi, t]$. 
These are governed by the Boltzmann equations, 
\begin{align}
\laq{BEBI} 
\frac{\partial f_\phi[p_\phi, t]}{\partial t} - p_\phi H \frac{\partial f_\phi[p_\phi, t]}{\partial p_\phi} & = C^\phi[p_\phi, t], \\
\frac{\partial f_\psi[p_\psi, t]}{\partial t} - p_\psi H \frac{\partial f_\psi[p_\psi, t]}{\partial p_\psi} & = C^\psi[p_\psi, t],
\end{align}
where $C^\phi$ and $C^\psi$ are the collision terms to be described in detail below. Following the motivation outlined above, we make the following ansatz for the phase space distribution functions:
\begin{align}
\laq{quasi}
f_\phi[p_\phi,t] & = \hat f_\phi[p_\phi,t] + f_\phi^{\rm th}[p_\phi,t], \\
f_\psi[p_\psi,t] & =  f_\psi^{\rm th}[p_\psi,t],
\end{align}
where 
\begin{align}
\hat f_\phi[p_\phi,t] & = (2\pi)^3 \hat n_\phi[t] \delta^3(\vec p_\phi) =  (2\pi)^3 \hat n_\phi[t] \left[\frac{\delta(p_\phi)}{4\pi p_\phi^2}\right], \\
f_\phi^{\rm th}[p_\phi,t] & = \left\{ \exp\left[ \frac{E_\phi - \mu_\phi[t]}{T[t]}  \right] -1  \right\}^{-1},  \\
f_\psi^{\rm th}[p_\psi,t] & = \left\{ \exp\left[ \frac{E_\psi - \mu_\psi[t]}{T[t]}  \right] +1  \right\}^{-1}, 
\end{align}
with $E_i = \sqrt{p_i^2+m_i^2}$ for $i = \phi, \psi$. 
In particular, we have separated the scalar phase-space distribution into condensate (denoted by the hat) and thermal components, and we have assumed the thermal components are well-described by the standard equilibrium forms with temperature $T[t]$ and chemical potential $\mu_i[t]$. We note that $\m_i[t]$  is not assumed to take any 
specific value associated with chemical/kinetic equilibrium. Rather, we will follow its dynamical evolution from generic initial conditions. As we will show below $\m_\f[t]=m_\f$ is the condition for the system to reach kinetic equilibrium. 

It will be convenient to follow the integrated quantities such as the number density, energy density, and pressure, defined as
\begin{align}
n_i[t] & = g_i \! \int \! \frac{d^3 p_i}{(2\pi)^3} \, f_i[p_i,t], \\
\rho_i[t] &  = g_i \! \int \! \frac{d^3 p_i}{(2\pi)^3} \, E_i \, f_i[p_i,t], \\
P_i[t] & = g_i \! \int \! \frac{d^3 p_i}{(2\pi)^3} \, \frac{p_i}{3E_i} \, f_i[p_i,t],
\end{align}
for $i = \phi, \psi$. Here $g_i$ counts the number of internal degrees of freedom for species $i$ ($g_\phi = 1$, $g_\psi = 2$).
In fact,  for the scalar $\phi$ we can isolate the condensate and thermal components by defining a type of projection operator $\hat P$ that excludes a small region around $p_\phi = 0$ in the $p_\phi$ integral. In particular, we define
\begin{align}
 \hat P \! \int \! \frac{d^3 p_\phi}{(2\pi)^3} & \equiv   \frac{1}{(2\pi)^3} \int d\Omega_\phi \int_\epsilon^\infty p_\phi^2 \, dp_\phi, \\
[1- \hat P] \! \int \! \frac{d^3 p_\phi}{(2\pi)^3} & \equiv   \frac{1}{(2\pi)^3} \int d\Omega_\phi \int_0^\epsilon p_\phi^2 \, dp_\phi,
\end{align}
where $\e$ is a small parameter.
We then have the relations 
 \begin{align}
[1- \hat P] \! \int \! \frac{d^3 p_\phi}{(2\pi)^3} \, f_\phi[p_\phi,t]  & = [1- \hat P] \! \int \! \frac{d^3 p_\phi}{(2\pi)^3} \, \hat f_\phi[p_\phi,t] = \hat n_\phi[t], \\
\hat P \! \int \! \frac{d^3 p_\phi}{(2\pi)^3} \, f_\phi[p_\phi,t]  & =  \hat P \! \int \! \frac{d^3 p_\phi}{(2\pi)^3} \,  f_\phi^{\rm th}[p_\phi,t] =  n_\phi^{\rm th}[t].
\end{align}

Using the projectors, the Boltzmann equations for the number densities can then be written as 
\begin{align}
\laq{fth}
 \dot {\hat n}_\phi + 3 H \hat n_\phi & = {\cal C}^{\phi_c},\\
\laq{ncon}
 \dot  n_\phi^{\rm th} + 3 H  n_\phi^{\rm th} & = {\cal C}^{\phi}, \\
\laq{psith}
 \dot {n}_\psi + 3 H  n_\psi & = {\cal C}^{\psi},
\end{align}
where the ${\cal C}$s on the r.h.s represent the corresponding integrated collision terms, which will be discussed below.  
Similarly, we can write the Boltzmann equations for the energy densities as 
\begin{align}
\laq{rhocond}
 \dot {\hat \rho}_\phi + 3 H   ( \hat \rho_\phi +  \hat P_\phi) & = {\cal E}^{\phi_c},\\
\laq{rhofth}
\dot  \rho_\phi^{\rm th} + 3 H ( \rho_\phi^{\rm th} +  P_\phi^{\rm th}    )  & = {\cal E}^{\phi}, \\
\laq{rhopsith}
\dot { \rho}_\psi + 3 H  ( \rho_\psi +  P_\psi  ) & = {\cal E}^{\psi},
\end{align}
where the ${\cal E}$s on the r.h.s represent the corresponding integrated collision terms. Summing the three equations above, we can write a Boltzmann equation for the total energy density:
\begin{align}
\laq{energycons}
 \dot { \rho}_{\rm tot} + 3 H ( \rho_{\rm tot} + P_{\rm tot} ) & = {\cal E}^{\phi_c} +  {\cal E}^{\phi} +  {\cal E}^{\psi} = 0, 
\end{align}
where $ \rho_{\rm tot} \equiv  \hat \rho_{\phi} +  \rho_{\phi}^{\rm th} +  \rho_{\psi}$,  
$P_{\rm tot} \equiv  \hat P_{\phi} +  P_{\phi}^{\rm th} +  P_{\psi}$,  and 
the last equality follows from energy conservation. In particular, for the condensate we have 
\begin{equation}
\hat \rho_\phi = m_\phi \hat n_\phi, ~~~~~~~ \hat P_\phi = 0. 
\end{equation}
On the other hand, the thermal $\f$ and $\p$ components are typically relativistic,  
leading to
\beq P_{\f}^{\rm th} \simeq \rho_\f^{\rm th}/3, ~~~~~~~  P_\p \simeq \rho_\p/3.\eeq  
We now turn to the collision terms for the various reactions. 

\subsubsection{Decay/inverse decay $\phi \leftrightarrow \psi\psi$}

We first consider the decay/inverse decay process 
\begin{equation}
\phi(p_\phi) \leftrightarrow \psi(p_1) + \psi(p_2).
\end{equation}
The unintegrated $\phi$ collision term $C^\phi_{\phi \leftrightarrow \psi\psi}$ is 
\begin{align}
C^\phi_{\phi \leftrightarrow \psi\psi} &= -\frac{1}{S_\psi} \frac{1}{g_\phi} \frac{1}{2E_\phi} \sum_{\rm spins} \int \! d\Pi_1 \, d\Pi_2 \,
(2\pi)^4 \delta^4(p_\phi-p_1-p_2) |{\cal M}_{\phi\rightarrow \psi \psi}|^2 \\
& \times \left\{  f_\phi[p_\phi] (1-  f_\psi[p_1] ) ( 1-  f_\psi[p_2] ) - (1+  f_\phi[p_\phi] )   f_\psi[p_1]   f_\psi[p_2]   \right\}. \nonumber
\end{align}
Here $1/S_\psi = 1/2$ is a symmetry factor accounting for identical $\psi$ particles,
$g_\phi = 1$, and 
\begin{align}
d\Pi_i \equiv \frac{d^3 p_i}{(2\pi)^3 2 E_i}. 
\end{align}

We now integrate over the $\phi$ phase space to derive the integrated collision terms, using the projection operators $[1-\hat P]$ and $\hat P$ to isolate the condensate and thermal components, respectively. For the condensate term, we obtain 
\begin{align}
{\cal C}^{\phi_c}_{\phi_c \leftrightarrow \psi\psi}  
& = g_\phi  [1-\hat P]  \!  \int \! \frac{d^3 p_\phi}{(2\pi)^3 } C^\phi_{\phi \leftrightarrow \psi\psi}  \\
&= -\frac{1}{S_\psi}  \sum_{\rm spins}  [1-\hat P] \!  \int \! d\Pi_\phi\, d\Pi_1 \, d\Pi_2  \,
(2\pi)^4 \delta^4(p_\phi-p_1-p_2) |{\cal M}_{\phi\rightarrow \psi\psi}|^2  
\nonumber \\ & \times   
\hat f_\phi[p_\phi] (1-  f_\psi[p_1] -  f_\psi[p_2] ) .   \nonumber
\end{align}
All integrals can be carried out analytically, leading to the simple expression
\begin{align}
\label{eq:C-phi-decay-C}
{\cal C}^{\phi_c}_{\phi_c \leftrightarrow \psi\psi}   = -\Gamma_{\phi\to \psi\psi} \left(  1- 2 f_\psi  [E_\psi = m_\phi/2]\right)  \hat n_\phi,
\end{align}
where $\Gamma_{\phi\to \psi\psi}$ is given in  \eqref{eq:phi-decay-width-1} (See Appendix~\ref{app:1} or Eq.~(\ref{eq:phi-decay-width}) for the mixed real and pseudo-scalar coupling case and full mass dependence). We note that the collision term \eq{C-phi-decay-C} reproduces the effective $\phi$ condensate decay rate advertised above in \Eq{decay}.

Next, for the thermal component we obtain 
\begin{align}
{\cal C}^\phi_{\phi \leftrightarrow \psi\psi}  
& = g_\phi  \hat P  \!  \int \! \frac{d^3 p_\phi}{(2\pi)^3 } C^\phi_{\phi \leftrightarrow \psi\psi}  \\
&= -\frac{1}{S_\psi}  \sum_{\rm spins}  \hat P \!  \int \!  d\Pi_\phi\, d\Pi_1 \, d\Pi_2 \,
(2\pi)^4 \delta^4(p_\phi-p_1-p_2) |{\cal M}_{\phi\rightarrow \psi\psi}|^2  
\nonumber \\ 
& \times \left\{  f^{\rm th}_\phi[p_\phi] (1-  f_\psi[p_1]) (1-  f_\psi[p_2]) - (1+  f^{\rm th}_\phi[p_\phi] )   f_\psi[p_1]   f_\psi[p_2]   \right\}. \nonumber
\end{align}
We can reduce the collision term to the following double integral expression:
\begin{align}
\label{eq:C-phi-decay-th}
{\cal C}^\phi_{\phi \leftrightarrow \psi\psi} &  = -\frac{m_\phi \Gamma_{\phi\to \psi\psi}}{2 \pi^2 \beta_\psi}
 \int_{m_\phi}^\infty\! d E_\phi \int_{E_1^-}^{E_1^+}\! d E_1  
\times \left\{  f^{\rm th}_\phi[p_\phi] (1-  f_\psi[p_1] -  f_\psi[p_2] ) -   f_\psi[p_1]   f_\psi[p_2]   \right\},
\end{align}
where the integration limits $E_1^{\pm}$ are  (see Appendix~\ref{app:2})
\begin{align}
E_1^\pm & = \frac{1}{2} (E_\phi \pm \beta_\psi p_\phi).
\end{align}
The phase space distributions here are regarded as functions of the energies $E_i = \sqrt{p_i^2 + m_i^2}$, and the energy conservation condition $E_\phi = E_1+E_2$ implies  
$f_\psi[p_2] = f_\psi[E_2 = E_\phi - E_1]$.

Following similar steps, the integrated collision term ${\cal C}^\psi_{\phi \leftrightarrow \psi\psi}$ can be related to the $\phi$ collision terms above as 
\begin{align}
{\cal C}^\psi_{\phi \leftrightarrow \psi\psi}  
=  - N_\psi  \, [ \,  {\cal C}^{\phi_c}_{\phi_c \leftrightarrow \psi\psi} +  {\cal C}^\phi_{\phi \leftrightarrow \psi\psi} \,], \nonumber
\end{align}
where the factor $N_\psi = 2$ is present because two $\psi$ particles are lost in the reaction. 

 We mention that ${\cal C}^{\phi_c}_{\phi_c \leftrightarrow \psi\psi}, -{\cal C}^{\psi }_{\phi_c \leftrightarrow \psi\psi} \propto e^{m_{\f}/2T}-e^{\m_{\p}/T}$ 
biases $\m_\p\to  m_\f/2$, at which point the DM condensate decay is Pauli blocked. Furthermore, 
 ${\cal C}^\phi_{\phi \leftrightarrow \psi\psi},-{\cal C}^\psi_{\phi \leftrightarrow \psi\psi} \propto e^{\m_\f/T}-e^{2\m_\p/T}$ biases $\m_\f\to 2\m_\p$. Therefore the chemical/kinetic equilibrium for decays/inverse decays  is achieved for $\mu_\psi = \mu_\f/2 =m_\f/2.$

\subsubsection{Annihilation/inverse annihilation $\phi \phi \leftrightarrow \psi\psi$}

Next, we consider the process 
\begin{equation}
\phi(p_\phi) + \phi(p_2) \leftrightarrow \psi(p_1) + \psi(p_2).
\end{equation}
The unintegrated $\phi$ collision term $C^\phi_{\phi \phi \leftrightarrow \psi\psi}$ is 
\begin{align}
C^\phi_{\phi \phi \leftrightarrow \psi\psi} &= -\frac{1}{S_\psi} \frac{1}{g_\phi} \frac{1}{2E_\phi}  \sum_{\rm spins} 
\int \! d\Pi_2 \, d\Pi_3 \, d\Pi_4 \,
(2\pi)^4 \delta^4(p_\phi + p_2-p_3 - p_4) |{\cal M}_{\phi \phi \rightarrow \psi \psi}|^2 ~~~~~~~~~~~~ \nonumber \\
& \times \left\{  f_\phi[p_\phi]  f_\phi[p_2] (1-  f_\psi[p_3]) (1-  f_\psi[p_4]) - (1+  f_\phi[p_\phi] )  (1+  f_\phi[p_2] )   f_\psi[p_3]   f_\psi[p_4]   \right\} .
\end{align}
Next, we integrate over the $\phi$ phase space to obtain the integrated collision terms, again employing the projection operators $[1-\hat P]$ and $\hat P$ to isolate the condensate and thermal components, respectively. We also include multiplicative factors $N_\phi = 2$  (two $\phi$ particles are lost) and a factor $1/S_\phi = 1/2$  (identical $\phi$ particles). There are two contributions to the condensate collision term. First, we have a term accounting for the annihilation of two condensate particles: 
\begin{align}
{\cal C}^{\phi_c}_{\phi_c \phi_c \leftrightarrow \psi\psi}  
& = \frac{N_\phi}{S_\phi} g_\phi  [1-\hat P]_\phi  \!  \int \! \frac{d^3 p_\phi}{(2\pi)^3 }   [1-\hat P]_2 C^\phi_{\phi \phi \leftrightarrow \psi\psi}  \\
&= -  \frac{N_\phi}{S_\phi S_\psi}    \sum_{\rm spins}  [1-\hat P]_\phi [1-\hat P]_2
\int \! d\Pi_\phi \, d\Pi_2 \, d\Pi_3 \, d\Pi_4 \nonumber \\ &
 \times (2\pi)^4 \delta^4(p_\phi + p_2-p_3 - p_4) |{\cal M}_{\phi \phi \rightarrow \psi \psi}|^2  \, 
    \hat f_\phi[p_\phi]  \hat f_\phi[p_2] (1-  f_\psi[p_3] -  f_\psi[p_4] ), \nonumber
\end{align}
where the subscript on the projector refers to the $\phi$ momentum. 
All integrals can be carried out analytically, leading to the simple expression
\begin{align}
\label{eq:C-phi-ann-cc}
{\cal C}^{\phi_c}_{\phi_c \phi_c \leftrightarrow \psi\psi}   = -  \frac{N_\phi}{S_\phi S_\psi} \frac{1}{32\pi m_\phi^2}  \sum_{\rm spins}   |{\cal M}_{\phi\phi\rightarrow \psi \psi}|^2  \bigg\vert_{\vec p_\phi = \vec p_2 = 0}
\left[1-\frac{m_\psi^2}{m_\phi^2}\right]^{1/2}
\!\! \left(  1- 2 f_\psi  [E_\psi = m_\phi]\right)  \hat n^2_\phi .
\end{align}

There is also a collision term accounting for the annihilation of a condensate scalar with a thermal one:
\begin{align}
\laq{22}
{\cal C}^{\phi_c}_{\phi_c \phi \leftrightarrow \psi\psi}  
& = \frac{N_\phi}{S_\phi} g_\phi  [1-\hat P]_\phi  \!  \int \! \frac{d^3 p_\phi}{(2\pi)^3 }   \hat P_2  \, C^\phi_{\phi \phi \leftrightarrow \psi\psi}  \\
&= -  \frac{N_\phi}{S_\phi S_\psi}    \sum_{\rm spins}  [1-\hat P]_\phi \, \hat P_2 \,
\int \! 
d\Pi_\phi \, d\Pi_2 \, d\Pi_3 \, d\Pi_4
 \, (2\pi)^4 \delta^4(p_\phi + p_2-p_3 - p_4)  \, 
\nonumber \\
& 
 \times |{\cal M}_{\phi\phi\rightarrow \psi \psi}|^2 \,   \hat f_\phi[p_\phi]  \left\{  f_\phi^{\rm th}[p_2] (1-  f_\psi[p_3] -  f_\psi[p_4] ) -  f_\psi[p_3] f_\psi[p_4] \right\},~~~~~\nonumber
\end{align}
We can reduce the collision term to the following double integral expression:
\begin{align}
\label{eq:C-phi-ann-cth}
{\cal C}^{\phi_c}_{\phi_c \phi \leftrightarrow \psi\psi} &  = -    \frac{N_\phi}{S_\phi S_\psi}   \frac{\hat n_\phi}{64 \pi^3 m_\phi}
 \sum_{\rm spins}   \int_{m_\phi}^\infty\! d E_2 \int_{E_3^-}^{E_3^+}\! d E_3  \,  |{\cal M}_{\phi\phi\rightarrow \psi \psi}|^2 
 \nonumber \\ & \times 
 \left\{  f_\phi^{\rm th}[p_2] (1-  f_\psi[p_3] -  f_\psi[p_4] )-  f_\psi[p_3] f_\psi[p_4]    \right\},
\end{align}
where the integration limits $E_3^{\pm}$ are given (see Appendix~\ref{app:2}):
\begin{align}
E_3^\pm & = \frac{1}{2} (E_2+m_\phi \pm \kappa_\psi \, p_2), ~~~~ \kappa_\psi = \left(1-\frac{2m_\psi^2}{m_\phi (E_2 + m_\phi)}\right)^{1/2}.
\end{align}
Again, the phase space distributions here are regarded as functions of the energies $E_i = \sqrt{p_i^2 + m_i^2}$, and the energy conservation condition $m_\phi+ E_2  = E_3+E_4$ implies  
$f_\psi[p_4] = f_\psi[E_4 = m_\phi + E_2 - E_3]$. 

Next, it is straightforward to see that the 
integrated collision term of the thermal $\phi$ component, which arises from the annihilation with a condensate scalar,
is identical to ${\cal C}^{\phi_c}_{\phi_c \phi \leftrightarrow \psi\psi}$ considered above:
\begin{align}
{\cal C}^{\phi}_{\phi \phi_c  \leftrightarrow \psi\psi}  = {\cal C}^{\phi_c}_{\phi_c \phi \leftrightarrow \psi\psi}.
\end{align}
We  also note that there is a collision term involving the annihilation of two thermal $\phi$ particles, ${\cal C}^{\phi}_{\phi \phi \leftrightarrow \psi\psi}$. We neglect the effect of this term since it is subdominant to  the other terms.

Finally, we consider the integrated $\psi$ collision term ${\cal C}^{\psi}_{\phi \phi \leftrightarrow \psi\psi}$, which can be related to the $\phi$ collision terms as follows:
\begin{align}
{\cal C}^{\psi}_{\phi \phi \leftrightarrow \psi\psi}  
&   = - \left[    {\cal C}^{\phi_c}_{\phi_c \phi_c \leftrightarrow \psi\psi}  +  {\cal C}^{\phi_c}_{\phi_c \phi \leftrightarrow \psi\psi}  +  {\cal C}^{\phi}_{\phi \phi_c \leftrightarrow \psi\psi}    +  {\cal C}^{\phi}_{\phi \phi \leftrightarrow \psi\psi}     \right]
\end{align}

We mention that   ${\cal C}^{\phi_c}_{\phi_c \f\leftrightarrow \psi\psi}, -{\cal C}^{\psi }_{\phi_c\f \leftrightarrow \psi\psi} \propto e^{(m_{\f}+\m_\f)/T}-e^{2\m_{\p}/T}$ biases $\m_\p\to  (\m_\f+m_\f)/2$, and  ${\cal C}^{\phi_c}_{\phi_c \f_c\leftrightarrow \psi\psi}, -{\cal C}^{\psi }_{\phi_c\f_c \leftrightarrow \psi\psi} \propto e^{m_{\f}/T}-e^{\m_{\p}/T}$ 
biases $\m_\p\to  \m_\f$. 
Therefore the chemical/kinetic equilibrium for the annihilation processes is realized when $\mu_\f=\mu_\p=m_\f.$

\subsubsection{Scattering $\phi \psi \leftrightarrow \phi\psi$}

Next, we consider the scattering process 
\begin{equation}
\phi(p_\phi) + \psi(p_2) \leftrightarrow \phi(p_3) + \psi(p_4).
\end{equation}
The unintegrated $\phi$ collision term $C^\phi_{\phi \psi \leftrightarrow \phi\psi}$ is 
\begin{align}
C^\phi_{\phi \psi \leftrightarrow \phi \psi} &= - \frac{1}{g_\phi} \frac{1}{2E_\phi}  \sum_{\rm spins} 
\int \! 
d\Pi_2 \, d\Pi_3  \, d\Pi_4 \,
(2\pi)^4 \delta^4(p_\phi + p_2-p_3 - p_4) |{\cal M}_{\phi \psi \rightarrow \phi \psi}|^2 ~~~~~~~~ \nonumber \\
& \times \left\{  f_\phi[p_\phi]  f_\psi[p_2] (1 + f_\phi[p_3] ) (1-  f_\psi[p_4] ) - (1+  f_\phi[p_\phi] )  (1-  f_\psi[p_2] )   f_\phi[p_3]   f_\psi[p_4]   \right\}.
\end{align}
Expanding out the phase space factors, we find several terms. The first one involves two $\phi$ condensate distributions, corresponding to the process $\phi_c \psi \leftrightarrow \phi_c\psi$. It is given by
\begin{equation}
\label{eq:scatter-CC}
\hat f_\phi[p_\phi] \hat f_\phi[p_3]   \left(  f_\psi[p_2] - f_\psi[p_4]  \right)    ~~~~~~~~~~ (\phi_c \psi \leftrightarrow \phi_c \psi).
\end{equation}
It is easily seen that this term vanishes since the delta functions from the condensate distributions enforce $\vec p_\phi = \vec p_3 = 0$, and therefore the energy conserving delta function enforces $E_2 = E_4$, implying  $\left(  f_\psi[p_2] - f_\psi[p_4]  \right) = 0$. Thus, Eq.~(\ref{eq:scatter-CC}) vanishes. 

The second class of terms encountered in the phase space factor involve one $\phi$ condensate and one thermal $\phi$, e.g., for the process $\phi_c \psi \leftrightarrow \phi\psi$ we find
\begin{equation}
\label{eq:scatter-Cth}
\hat f_\phi[p_\phi] \left\{  f_\phi^{\rm th}[p_3]   \left(  f_\psi[p_2] - f_\psi[p_4]  \right)     +  f_\psi[p_2]   \left( 1- f_\psi[p_4]\right)     \right\}   ~~~~~~~~~ (\phi_c \psi \leftrightarrow \phi \psi),
\end{equation}
and an analogous expression for the process $\phi \psi \leftrightarrow \phi_c \psi$. Plugging in the explicit forms of the equilibrium distributions and using energy conservation, we find 
\begin{equation}
\label{eq:scatter-Cth-prop}
\hat f_\phi[p_\phi] \left\{  f_\phi^{\rm th}[p_3]   \left(  f_\psi[p_2] - f_\psi[p_4]  \right)     +  f_\psi[p_2]   \left( 1- f_\psi[p_4]\right)     \right\}    \propto  e^{(E_2+\mu_\phi)/T} - e^{(E_2+m_\phi)/T}.
\end{equation}
Notice that once $\mu_\phi = m_\phi$ is established, this factor also vanishes. 
However the relevant collision term is always slower than the other collision terms, and we  
therefore neglect it in the following. 

Finally, there is a term corresponding to the scattering involving two thermal $\phi$ particles, $\phi \psi \leftrightarrow \phi\psi$. 
This contribution vanishes for our ansatz.
{Motivated by the equilibrium of this number-conserving process, 
we consider kinetic equilibrium of the system to be achieved when the condition $\mu_\phi[t]=m_\f$ is satisfied.

\section{DM stability with exotic fermions}
\lac{DMODS}
In this section, we study the stability of the  $\phi$ DM condensate against decay and scattering processes in a simple model with a new exotic fermion $\psi$.
Let us first describe the approach to thermalization in this scenario. Suppose that after the $\phi$ starts oscillating,
 in addition to the condensate, there is plasma with number density $\O(m_\phi^3)$ and a typical particle momentum of $p_{\rm typ}\sim m_\phi$. As will be discussed below in Sec.~\ref{secpr}, this is expected to arise at early times due to broad parametric resonance production of $\psi$.  
It is known that for $p_{\rm typ}\gtrsim m_\phi$ in such a system, particle decays  lead to the momenta being fully randomized on the timescale of $\sim \(\frac{ y^2 m_\phi}{16\pi } (\frac{m_\phi}{p_{\rm typ}})^n\)^{-1}$, with positive power $n$; see, e.g., Refs.~\cite{Hannestad:2005ex,Archidiacono:2013dua,Basboll:2008fx,Escudero:2019gfk,Barenboim:2020vrr} for investigations of such anisotropic plasma as well as further discussion below in \Sec{CnuB-free-stream}. 
Therefore, at the beginning stages, the thermalization rate is parametrically $\sim \frac{ y^2 m_\phi}{16\pi } $,
which is model-independent. 
Thus the distribution functions for the plasma components follow the quasi-equilibrium forms~\cite{Barenboim:2020vrr}, justifying the ansatz \Eq{quasi} and the theoretical setup described in the previous section, at least for timescales longer than $(\frac{ y^2 m_\phi}{16\pi })^{-1}$. At this timescale, the system consists of the nearly homogeneous and isotropic dark radiation plasma and the DM  condensate, which then evolve according to the Boltzmann equations from the previous section.

\subsection{Case with chirality suppression}
\lac{cs}
We first consider the stability of the $\phi$ condensate, the emergence of the thermalized $\phi$, $\psi$ dark radiation component, and its phenomenological implications in the case that the self-annihilation of the $\phi$ condensate is chirality suppressed. 
We therefore take the limit $m_\psi\to 0$ in this subsection.\footnote{We note that qualitatively similar results will apply for the case of a purely pseudoscalar coupling, in which case the $\phi$ annihilation is velocity suppressed; see Appendix~\ref{app:1} for further details.}
In the presence of the chirality suppression, the annihilation of $\f$ is dominated by high momentum modes with $p_\f \gtrsim m_\f. $ 
In particular, the annihilation between two condensate particles, $\f_c\f_c \to  \psi \psi $ is negligible, i.e. ${\cal C}^{\f_c,\p}_{\f_c \f_c\to \p\p}\rightarrow 0$. For simplicity, we neglect the effect of the Hubble expansion in this subsection, $H = 0$.

The numerical solution to the Boltzmann equations,
\Eqs{fth}, \eq{ncon}, \eq{psith} and \eq{energycons},  are presented in Fig.~\ref{fig:boltz}. 
The initial conditions are taken to be $n_\f[0]/m_\f^3=10^{8}, T[0]=3/2m_\f, \mu_\psi[0]=\mu_\f[0]=0$. We fix $y=10^{-8}$ and neglect the Hubble expansion.
We take account of the annihilation and decay collision terms, ${\cal C}_{\f \f_c\leftrightarrow \p\p}^{\f,\f_c,\p}, {\cal C}_{\f/\f_c \leftrightarrow \p\p}^{\f,\f_c,\p}$, respectively, and neglect the others, which are sub-leading. For the relevant matrix element expressions, see Appendix~\ref{app:1}. 
To study the late time evolution of the DM condensate and the dark sector thermalization, we expand $\m_\f$  and $\m_\psi$ around $0$ 
in the collision terms of ${\cal C}^{\f_c,\f,\p}_{\f_c \f \leftrightarrow \psi \psi}$ and ${\cal C}^{\f,\p}_{\f \leftrightarrow \psi \psi}$. 
This is done up to the third power in both $\m_\f \AND \m_\psi$, which is a good approximation in the regime $T\gtrsim \m_{\rm \psi}, \m_\f.$
This Taylor expansion is also carried out for $n_\f^{\rm th}, \rho_\f^{\rm th}, n_\psi,\AND \rho_\psi$. 
On the other hand, we use the full expression for ${\cal C}^{\f_c,\p}_{\f_c\leftrightarrow \p\p}$ as it can be evaluated exactly analytically. 
\begin{figure}[t!]
\begin{center}  
\includegraphics[width=135mm]{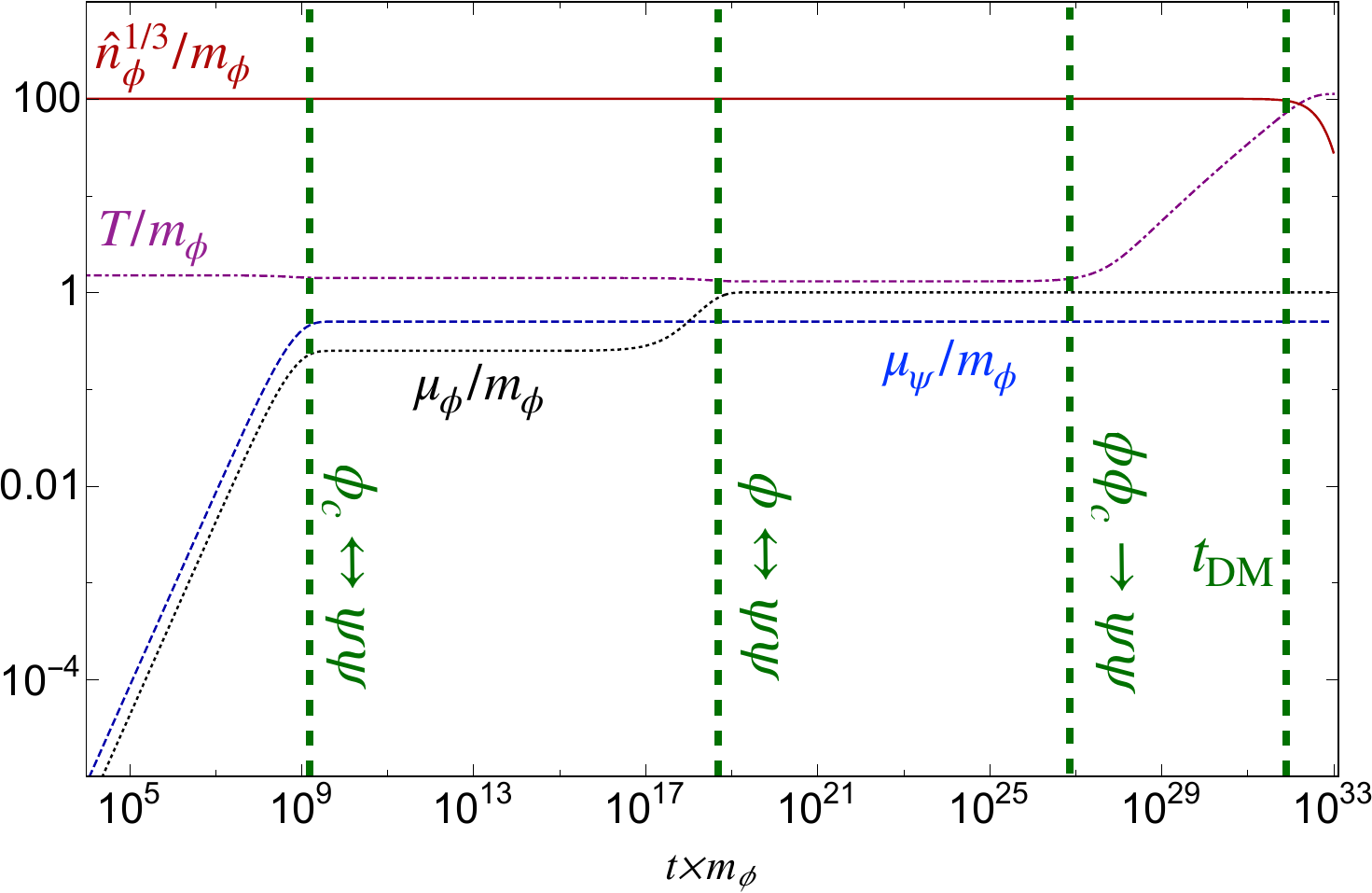}
\end{center}
\caption{Stability and thermalization in the model of scalar DM $\phi$ interacting with exotic fermions $\psi$, with $m_\psi=0$. We take the initial conditions $\hat n_\f[0]/m_\f^3=10^{8}, T[0]=3/2m_\f, \mu_\psi[0]=\mu_\f[0]=0$ and fix  $y=10^{-8}$. We also show relevant timescales corresponding to various interactions in vertical green dashed lines (see the main text for details). 
} 
\label{fig:boltz}
\end{figure}

Let us now describe the evolution of the system using semi-analytic arguments. Given the ansatz of the initial distribution, 
the decay of the $\phi$ condensate rapidly produces $\psi$ particles with momenta $m_\phi/2$. 
This process is quickly saturated due to Pauli blocking, $\mu_\psi \rightarrow m_\phi/2$, as indicated by the vertical green dashed line labeled $\phi_c \leftrightarrow \psi \psi$ in Fig.~\ref{fig:boltz}. 
The system subsequently approaches kinetic equilibrium 
 defined at a rate of\footnote{Strictly speaking, 
the collision term \eqref{eq:C-phi-decay-th} with our ansatz gives the r.h.s to be $-C_{\f \leftrightarrow \p\p}^\f/n_\f^{\rm th}\sim  {m_\f \G_\f}  \frac{(2\mu_\p -\mu_\f)}{T^2}$, with $n_\f^{\rm th}\sim T^3/\pi^2.$  
} 
\beq
\label{eq:Gamma-kin}
\G_{\rm kin} \equiv \frac{y^2 m^2_\f}{16\pi T}\,.
\eeq 
This represents the typical 
interaction rate of the splitting process of the ambient plasma  $\f \leftrightarrow \psi \psi$.  
The typical momentum of the fermion is taken to be of order $T$, and as such the Pauli blocking effect is not important and is thus omitted in the rate estimate, Eq.~(\ref{eq:Gamma-kin}). 
Although realistically, the evolution for timescales smaller than $1/\G_{\rm kin}|_{T\sim m_\phi}$ depends in detail on the initial distribution, 
 as we have explained, when the timescale is longer than $1/\G_{\rm kin}|_{T\sim m_\phi}$, the evolution  of bulk quantities should not depend much on the initial distribution with the number density and energy density fixed. 
Thus our estimates should be robust for timescales longer than $1/\G_{\rm kin}|_{T\sim m_\phi}$.
As a consistency check we have verified numerically that the late time evolution seen in Fig.~\ref{fig:boltz} is obtained for a variety of initial conditions for $\m_\p,\m_\f,T$.

Therefore, we expect that Fig.~\ref{fig:boltz} provides a good description of the evolution of the system to the right of the vertical green dashed line labeled $\f\leftrightarrow \p\p$.
One observes from the figure that $1/\G_{\rm kin}|_{T\sim m_\f}$, which corresponds to the conventional DM lifetime in vacuum, does not represent the lifetime of the DM condensate in our scenario, a consequence of Pauli blocking.

Next, the temperature of the plasma starts to grow at the timescale\footnote{More precisely, the timescale for the onset of temperature growth, i.e., when $\Delta T / T \sim {\cal O}(1)$, can be estimated from \Eq{groT} and is in agreement with the timescale \Eq{scat} for $T \sim m_\phi$.}   
\beq
\laq{scat}
\D t\sim  \G_{\rm scat}^{-1}\equiv \( \frac{y^4\hat{n}_\f}{32\pi T m_\f}\)^{-1},
\eeq
corresponding to the inverse of the characteristic
interaction rate of the  $\f \f_c\to \p\p$ scattering process, as indicated by the 
green vertical line labeled $\f \f_c\to \p\p$ in Fig.~\ref{fig:boltz}. 
The reason 
the temperature does not grow for $t\ll \D t$ is the number conservation associated with the rapid $1\to 2$ process, 
corresponding to the chemical equilibrium condition
\beq
\laq{decaycond}
\mu_\f=2\m_\p=m_\f \, .
\eeq 
At early times, while the $2\leftrightarrow2$ process is inefficient, the temperature does not grow because  number conservation and energy conservation fix the temperature. This is seen in Fig.~\ref{fig:boltz} to the
left of the $\f \f_c\to \p\p$ line.  
We also note that $\D t \gg 1/\G_{\rm kin}$ is typically satisfied as a consequence of 
 \Eq{con1}.
 Thus \Eq{decaycond} is satisfied in the whole range right to $\f\leftrightarrow \p\p$ line including the $t\gg \D t$ region.

When $t \gtrsim \D t$, the temperature starts to grow. 
This is because both the decay and annihilation processes are efficient, and the particle number is no longer conserved. 
This contradicts \Eq{decaycond}. Thus 
the condensate starts to evaporate.
Since the chemical potential is fixed to be \eq{decaycond} by the fastest reaction, the evaporation is partially Pauli blocked. 
Indeed the evaporation happens at the timescale of
\beq
\G_{\rm th}\sim \frac{1}{\D t} \frac{m_\f}{T}.
\eeq
The suppression by $m_\f/T$ is because the $2\to 2$ annihilation would be blocked (or in kinetic equilibrium) if $\m_\psi =\m_\f=m_\f$ were exact.  
Thus the collision term of the annihilation, i.e., \Eq{22}, is suppressed by $\O((\m_\psi -m_\f)/T,(\m_\f -m_\f)/T)\sim m_\f /T$ from \Eq{decaycond}.

The plasma temperature growth can be studied by solving the following Boltzmann equation,\footnote{More precisely, one can estimate the r.h.s from \eqref{eq:C-phi-ann-cth} to be $C_{\phi_c \f \leftrightarrow \p\p}^{\f_c} m_
\f \sim \frac{T^3}{\pi^2}[\frac{y^4 \hat{n}_\f}{32\pi Tm_\phi} \frac{m_\phi}{T}]
\log{\frac{T}{m_\phi}}\times m_\phi $.
}
\beq
\laq{groT}
 \dt{\(\rho_\f^{\rm th}+ \rho_\p\)} = (g_\f+7/8g_\p) \frac{\pi^2}{30}\dt{T^4}
 \sim \G_{\rm th} n_\phi^{\rm th} m_\f ,
\eeq
with $n_\f^{\rm th}\sim T^3/\pi^2.$ The last term is understood as the number changing rate per unit volume via $\f\f_c\to \p\p$ reaction 
by multiplying the transferred energy, $m_\f$, per each collision.  
This can be also checked from the behavior of the collision term of \Eq{C-phi-ann-cth}.
Integrating \Eq{groT} until 
 $ (g_\f+7/8g_\p) \frac{\pi^2}{30}{T^4[t_{\rm DM}]} \sim \hat n_\f[0] m_\f$ allows us to estimate the lifetime of  $\f$ condensate: 
 \beq
\laq{ltcs}
t_{\rm DM}\sim  \left[(g_\phi + \tfrac{7}{8}g_\psi) \frac{\pi^2}{30}\right]^{1/4}\frac{128 \pi^3}{3 y^4 \(m_\f \hat{n}_\f\)^{1/4}}
\sim\frac{1300}{y^4 \(m_\f \hat{n}_\f\)^{1/4}},
 \eeq
 where we have neglected the change of 
$\hat{n}_\f $ and taken $g_\f =1, g_\p=2.$
The timescale agrees well with the numerical estimation  for the benchmark in Fig.~\ref{fig:boltz}.

By requiring the lifetime to be  longer than the age of the Universe, $13.8$\,Gyr, and by assuming $\hat n_\f \simeq \rho_{\rm DM}/m_\f$ with $\rho_{\rm DM}$ being the current (global) DM density, 
we obtain the condition for $\f$ as the DM, 
\beq
\laq{uppery}
|y|\lesssim 2\times 10^{-7}.
\eeq 
Strictly speaking, we need to take account of the redshift in the estimation, which would not change the bound much. 
The inclusion of the Hubble expansion will be discussed next in \Sec{massive}.  

We can also estimate the dark radiation at the recombination era, $ z_{\rm rec}\sim 1100$, from 
\beq
\frac{(g_\phi+ \tfrac{7}{8} g_\psi)\pi^2}{30} T^4 \sim \(\frac{ \G_{\rm th} m_\phi n_\phi^{\rm th}}{H} \sim \rho_{\rm DM} \left. \frac{y^4 T}{32 \pi^3 H }\)\right|_{z=z_{\rm rec}}.
\eeq

Applying the constraint from \Eq{uppery}, we obtain
\beq
\D N_{\rm eff}[z_{\rm rec}] \lesssim  0.03    \times  \left(\frac{y}{2 \times 10^{-7}}\right)^{16/3}.
\eeq
Here 
\beq 
\D N_{\rm eff}[z]\equiv \frac{4}{7} \frac{30 \rho_{\rm  DS}[z]}{\pi^2 T_\nu[z]^4},
\eeq 
 with $T_\nu$ being the neutrino temperature. 

Thus, the amount of dark radiation produced is relatively small,
and the thermal components of the dark sector are unlikely to impact the evolution of the cosmic history given that $\f$ lives much longer than the age of the Universe. This is the case for $Q\ll 1$\footnote{When $Q\lesssim 1$ but it is not very small, there may be a similar effect 
as in the chirality unsuppressed case
 $m_\psi \neq 0$
 studied in \Sec{massive} since there is an induced effective fermion mass of $m_{\psi}^{\rm eff}\sim \sqrt{Q} m_\f \sim y \sqrt{\hat n_\phi/m_\f }\lesssim m_\f.$ One can find the effective mass by taking the time average of $(y \phi[t])^2$, which is identified as $(m_{\psi}^{ \rm eff})^2$. This may provide an efficient dark radiation 
 production mechanism soon before the recombination. 
} and $m_\psi=0$ (or with $m_\psi \neq 0$ and a $\phi \bar\psi \gamma^5 \psi$ coupling, see Appendix~\ref{app:1}). We will see in the next subsection that a qualitatively different situation arises for $m_\psi\neq 0$, where the evolution of energy densities can be quite different from that in the $\Lambda$CDM. 

\subsection{Case without chirality suppression}
\lac{massive}
Let us now turn on the mass term of the fermion, 
\beq
\d{\cal L}\supset -\frac{m_{\psi}}{2} \bar{\psi} \psi,
\eeq
with $m_\psi <m_\f/2$ being the aforementioned mass parameter of the fermion. 
The main difference from the massless case is that the $\f$ 
condensate 
can self-annihilate, $\f_c\f_c \to \p\p$.
Then we have the collision term from \Eq{C-phi-ann-cc}, 
\beq
{\cal C}^{\f_c}_{\f_c \f_c \leftrightarrow \psi\psi} \simeq 
-\frac{2y^4  m^2_\psi }{\pi m_\f^4 } \hat{n}_\f[t]^2 (1 - 2 f_\psi[m_\f]),
\eeq
where we have taken the leading term in the Taylor series in powers of $m_\psi/m_\phi$. 

The numerical result for the evolution of $\mu_i, T, \hat{n}_\phi$, including the collision term \eq{C-phi-ann-cc}, is presented in Fig.~\ref{fig:boltz2} where we take $m_\psi=m_\f/50$.  
We do not expand ${\cal C}^{\f_c,\p}_{\f_c \f_c \leftrightarrow \p\p}$ in terms of $T,\m_\f,\m_\p$  as it may be evaluated exactly analytically. 
The other parameters are the same as in Fig.\ref{fig:boltz}.

The time to the onset of temperature growth is shorter than in the chirality suppressed case due to the enhanced condensate annihilation. 
However, the duration of the temperature growth 
is longer than in the chirality suppressed case. This is due to 
the Pauli blocking factor $1-2f_\psi[m_\f]\sim m_\f/4T$ with $\m_\psi=m_\f/2$, which leads to a decrease in the annihilation rate as $T$ increases. 
In contrast, in the previous chirality suppressed case, $-\dt \log \hat{n}_\f \propto T$ increases when $T$ increases. 
By including the contributions from both $\f_c\f_c \to \p\p$ and $\f_c \f\to \p\p$ (see \Eq{ltcs}), the lifetime of the DM can be estimated as  
\beq 
t_{\rm DM}^{-1} \simeq \max{[0.2\frac{y^4 m^2_\psi }{  m_\f^4 } {\(m_\f \hat{n}_\f\)^{3/4}} ,~~ 0.0008 y^4 \(m_\f \hat{n}_\f\)^{1/4}]}
\eeq
where we have solved $  \dt{\(\rho_\f^{\rm th}+ \rho_\p\)} 
 \sim {\cal C}^{\f_c}_{\f_c \f_c} \times m_\f $ for the first component and again defined the lifetime from the condition 
$\(\rho_\f^{\rm th}+ \rho_\p\)= m_\f \hat{n}_\f[0]$ and neglected the decrease of  $\hat{n}_\f$ in the estimation.

This form agrees well with the numerical solution to the Boltzmann equations presented in Fig.~\ref{fig:boltz2}. When  $\f_c\f_c\to \p\p$  is the dominant scattering process, the 
constraint, $t_{\rm DM}> 13.8$\,Gyr, 
turns out to be 
\beq
\frac{m_{\psi}}{m_\f} \lesssim 0.8\(\frac{10^{-9}}{y} \)^2 \left(\frac{m_\f}{10^{-6}\EV}\right).
\eeq
A more stringent bound will be obtained in the following.

\begin{figure}[t!]
\begin{center}  
\includegraphics[width=135mm]{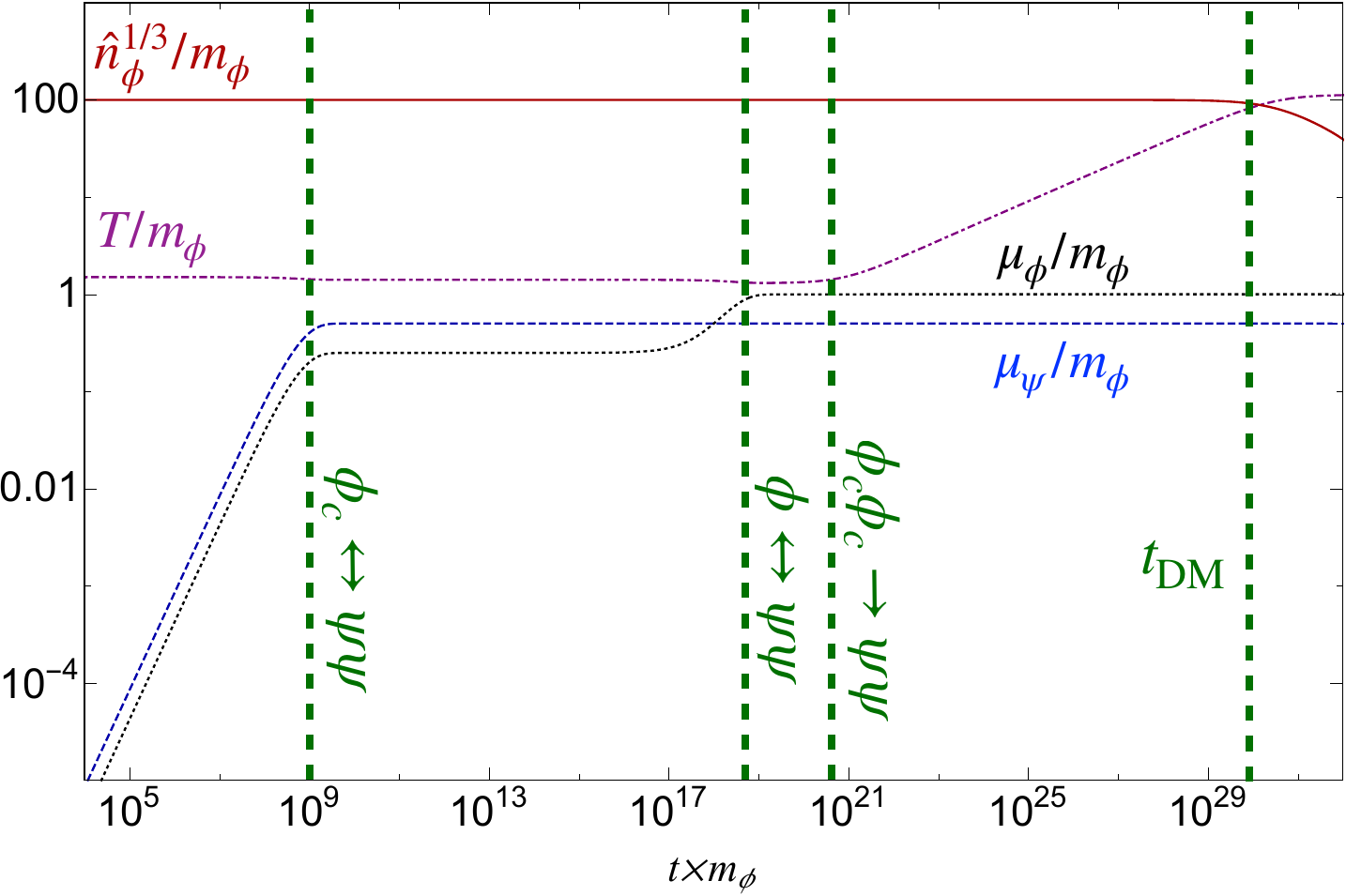}
\end{center}
\caption{Same as Fig.\ref{fig:boltz} except that we take $m_\psi=m_\f/50$. 
} 
\label{fig:boltz2}
\end{figure}

Intriguingly, the slow thermalization 
provides a production mechanism for abundant dark radiation in the early Universe. 
The dark radiation contribution can 
lead to slight deviations from the $\L$CDM predictions, while maintaining the cosmic stability of the DM condensate.
To study this quantitatively,
let us consider the Boltzmann equations including the effect 
of the Hubble expansion. 
To simplify the discussion let us focus on timescales longer than the time to reach kinetic equilibrium (right of the green dashed line labeled $\f \leftrightarrow \p\p$ in Fig. \ref{fig:boltz2}), i.e. $\G_{\rm kin}>H$, and 
let us first focus on the evolution when 
$
T\gg m_\f
$
is satisfied 
and the $\f_c \f_c \to \p\p$ contribution to scattering dominates. 
Then \Eq{decaycond} is satisfied.

The integrated Boltzmann equation for the $\f$ condensate number density is obtained from \Eqs{fth} and \eq{decaycond} as  
\beq
\laq{nphi}
\dot{\hat{n}}_\f+3H \hat{n}_\f\simeq - \frac{2y^4 m_\psi^2\hat{n}^2_\f }{\pi m_\f^4} (1-2f_\p[m_\f])\simeq - \frac{y^4 m_\psi^2 \hat{n}_\f^2}{2\pi m^3_\f T},
\eeq
while that for the 
thermal component of dark sector energy density is found from \Eqs{rhofth},  \eq{rhopsith}, and \eq{decaycond} to be
\beq
\laq{rad} \dot{\rho}_{\rm DS}+4H\rho_{\rm DS}\simeq  \frac{y^4 m_{\psi}^2\hat{n}_\f^2}{2 \pi m_\f^2 T}.
\eeq
Note that \eq{rad} also follows straightforwardly from energy conservation. 
Namely, the decrease of a $\f$ particle in the condensate corresponds to the increase of the radiation energy by $m_\f$.  
Here $\rho_{\rm DS}\simeq (g_\f + \tfrac{7}{8}g_\p) \pi^2 T^4/30=({11}/{4} ) \pi^2 T^4/30$ is the dark radiation of relativistic $\f$ and $\p$ (with $T\gg m_\f$). 
Although $\f_c \f_c\to \p\p$ is the reaction that transfers energy from the condensate to the fermions, the 
$\f, \p$ radiation readily shares the transferred energy since the system is in
kinetic equilibrium.
The r.h.s of both equations are suppressed by $m_\phi/T$ 
thanks to the Pauli exclusion principle.

By solving the equations one can estimate the final temperature of radiation at different epochs. 
Before showing the numerical result, we will present an approximate analytic calculation of
the evolution of the dark radiation and the resulting constraints. 
In our analytical estimate, we will neglect the decrease of $\hat n_\f a^3$ with time.

The radiation generated in a 
Hubble time $1/H$ can be estimated from \Eq{rad} and is given by $\delta \rho_{\rm DS} a^4 \sim \frac{y^4 m_{\psi}^2\hat{n}_\f^2}{2 \pi m_\f^2 T H}a^4$, with $a=1/(1+z)$ being the scale factor. 
By noting $\hat n_\f\propto a^{-3}$, we see that $a^4 \rho_{\rm DS}$ is dominantly generated at 
late times during the radiation- and matter-dominated eras.   
During these eras, we can therefore obtain an estimate of the dark radiation temperature from the equality $ H\rho_{\rm DS}\sim  \frac{y^4 m_\psi^2 \hat{n}_\f^2}{ 2\pi m_\f^2 T}$: 
\beq 
\laq{T}
T \sim \left.\(\frac{15 y^4 m_{\psi}^2 {\rho}_\f^2}{ (g_\f+\frac{7}{8}  g_\p)  \pi^3 H m_\f^4 } \)^{1/5}\right|_{z\gg z_{\rm DE}},
\eeq
where $z_{\rm DE}\simeq 0.3$ represents the redshift at the dark energy-matter equality.
This temperature scales as $a^{-4/5}\AND a^{-9/10}$ during the radiation- and matter-dominated eras, respectively.
Although the production of the dark radiation is most efficient at late times during these eras, $\rho_{\rm DS}/\rho_\f$ is largest at early times in the radiation-dominated era.\footnote{The production of $\rho_{\rm DS}/\rho_\f$ is dominant at the early stage, say the onset of oscillation of $\f$. 
However at that epoch we will have $Q\gg 1$ and the discussion here breaks down (see Sec.\,\ref{secpr} for the discussion in this regime.)  }
Also, the dark radiation scales $\rho_{\rm DS} a^4 \propto a^{4/5} \AND a^{2/5}$ in the radiation- and matter-dominated eras, respectively.

When dark energy is dominant, $z<z_{\rm DE}$, the produced $\d \rho_{\rm DS} a^4$ is smaller than that produced at $z=z_{\rm DE}$. 
Thus $\rho_{\rm DS}$ for $z<z_{\rm DE}$ can be approximated from the redshift of $\rho_{\rm DS}$ at $z=z_{\rm DE}$.
Using \Eq{T} and accounting for the redshift,  we obtain the dark sector temperature at the present epoch:
\beq
\laq{DRTt}
T[z=0]\sim \left. (1+z_{\rm DE})^{-1}\(\frac{15 y^4 m_\psi^2 {\rho}_\f^2}{ (\frac{7}{8}  g_\p +g_\f)  \pi^3 H m_\f^4 } \)^{1/5}\right|_{ z=z_{\rm DE}}.
\eeq
By using the DM density scaled from today,  $\rho_\f=\rho_{\rm DM}^{\rm \L CDM}\equiv(1+z)^3\rho_{\rm DM}$ and the $\L$CDM Hubble parameter
 we obtain $T[z]$ and thus the dark radiation with a given set of $y$ and $m_\f$. 
Interestingly, the resulting dark radiation does not depend on the initial conditions.

In order to satisfy the bounds from the cosmic microwave background (CMB) and baryon acoustic oscillation (BAO) observations~\cite{Aghanim:2018eyx,Blinov:2020hmc}, we need 
$\D N_{\rm eff}[z_{\rm rec}] \lesssim 1$.\footnote{Here we use a relaxed bound rather than $\D N_{\rm eff}\lesssim 0.3$\cite{Aghanim:2018eyx} since our conclusions here are based on order-of-magnitude estimates and since the dark radiation may not be 
free-streaming as we will discuss (see also, e.g., Ref.~\cite{Blinov:2020hmc} for discussion on the relaxed bound for non-free-streaming dark radiation).}
 This constraint turns out to be 
\begin{align}
\laq{dradi}
y\sqrt{\frac{m_\psi}{m_\f}}\lesssim 7\times 10^{-11}\sqrt{\frac{m_\f}{10^{-6}\EV}}  \text{  if  } T\gg m_\f
\text{  at  }z= z_{\rm rec}.
\end{align}
On the other hand, $\D N_{\rm eff}[z_{\rm rec}]= \O(0.01\text{-}1)$ may be tested in the future CMB and BAO observations~\cite{Kogut:2011xw, Abazajian:2016yjj, Baumann:2017lmt}.

We solve the integrated Boltzmann equations \eq{nphi} and \eq{rad} numerically for the benchmark  $y=10^{-9}$, $m_\f=10^{-4}\EV$, and $m_\psi=0.2 m_\f.$
The evolution of $\D N_{\rm eff}$ and $1-\rho_\f/\rho_{\rm DM}^{\rm \L CDM}$, i.e., the deviation of the DM density from the $\L$CDM prediction as a function of the redshift, is shown in Fig.~\ref{fig:DR} by the red dashed and solid lines, respectively. 
The initial condition is taken to be $\rho_\f[z_{\rm ini}]=\rho^{\L \rm CDM}_{\rm DM}$ and $T[z_{\rm ini}]=m_\f$ at $z_{\rm ini}=10^5$.\footnote{We take $z=10^5$ as the initial condition because we would like to show the evolution after $z=10^4-10^5$ at which the broad parametric resonance ends (see the next subsection). 
} Here 
$\rho^{\L \rm CDM}_{\rm DM}$ is the DM density in the $\L$CDM model. The temperature $T[z\ll z_{\rm ini}]$ does not depend much on the initial condition given that the DM dominates.
The resulting $\D N_{\rm eff}[z=z_{\rm rec}]\simeq 0.33$ for this benchmark. 
For comparison, we also consider the predictions of conventional decaying DM models (i.e., negligible effects from Pauli blocking). We show the decaying DM predictions for the same quantities  by the black lines in Fig.~\ref{fig:DR}, taking the same initial conditions and fixing the decay rate to $\G_{\rm Decaying~DM}=0.1 H_0$.\footnote{One may consider exactly the same Lagrangian \Eq{int} as a model of ordinary decaying DM. This is realized for large enough $m_\f$ 
so that the Pauli blocking
is irrelevant and small enough $y$ so 
that the decay rate in vacuum is suppressed. In other words, our findings may be considered as a different parameter region of the models of the decaying DM into fermions.} 
We note that the decaying DM case is close to the bound on the decay width given in \cite{Bringmann:2018jpr, Enqvist:2019tsa}. 
In this case, the DM density deviates by $\O(10)\%$ from $\L$CDM at the present epoch while the amount of dark radiation is negligible (significant) at $z\sim z_{\rm rec}$ ($z\sim 0$). 
In contrast, in our scenario, the DM density differs by $\O(1-10)\%$ from the $\L$CDM model today while having substantial dark radiation near recombination. 
Thus we expect that it is safer than the decaying DM with $\G_{\rm Decaying~DM}=0.1H_0.$
On the other hand, the production of dark radiation in our scenario is slower than in the decaying DM scenario at the earlier times while it ceases at late times 
$z\lesssim 1$. Thus the numerical solution confirms the previous discussions. Indeed, an analytic relation in our scenario can be obtained  from the redshift dependence (see \Eq{DRTt}),
\beq \laq{Neff}\frac{\D N_{\rm eff}[z=0]}{\D N_{\rm eff}[z_{\rm rec}]}= 
\left(\frac{H(z_{\rm rec}) (1+z_{\rm DE})}{H(z_{\rm DE})(1+z_{\rm rec})}\right)^{4/5}
= \O(10) .\eeq 
An analytic estimation gives this factor around 10, but numerically solving the evolution equation gives around $20$. 
This is quite different from the production of the dark radiation in the ordinary decaying DM scenario.
This slower DM ``decay" and larger $\D N_{\rm eff}[z_{\rm rec}]$ may help to alleviate the Hubble tensions, as will be discussed  in Sec.\,\ref{chap:Htension}.

\begin{figure}[t!]
\begin{center}  
\includegraphics[width=145mm]{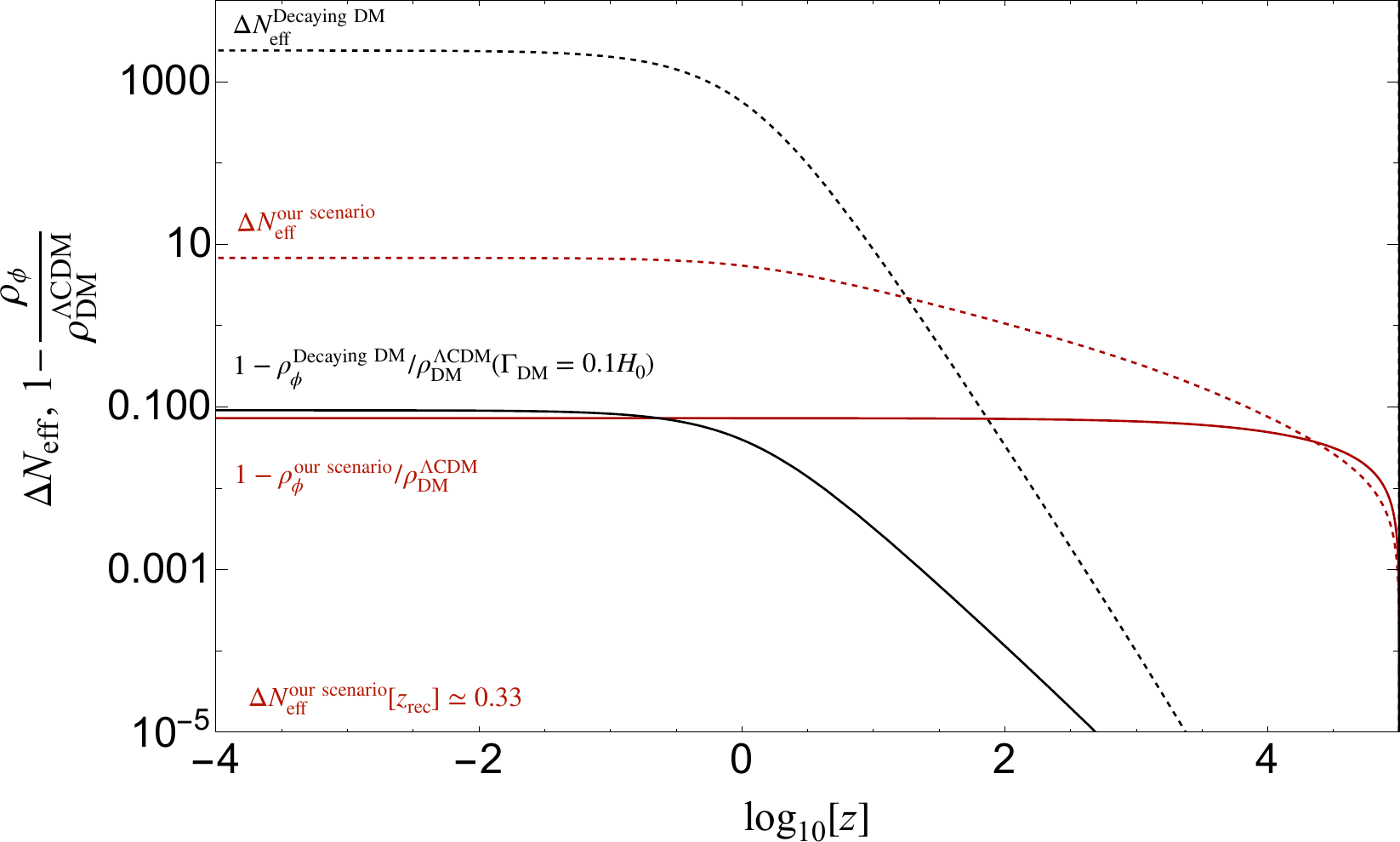}
\end{center}
\caption{ The evolution of $\D N_{\rm eff}[z]$ and $1-\rho_\f/\rho_{\rm DM}^{\rm \L CDM}$ (i.e., the deviation of the DM density from the $\L$CDM prediction) with redshift in our scenario (red dashed and solid lines, respectively).
The initial condition is set by $\rho_\f[z=10^5]=\rho^{\L \rm CDM}_{\rm DM},T[z=10^5]=m_\f$, though the result 
is relatively insensitive to the 
dark sector temperature.
We have fixed $y= 10^{-9}$, $m_\f=10^{-4}\EV$, and $m_\psi= m_\f/5.$
We find $\D N_{\rm eff}[z_{\rm rec}]\sim 0.33.$ 
For comparison, we also show the prediction for conventional decaying DM with decay rate $\G_{\rm Decaying~DM}=0.1 H_0$  and the same initial conditions (black lines). 
} 
\label{fig:DR}
\end{figure}

When $T\gg m_\f$ is not satisfied at some epoch, we obtain the stable DM with the Fermi sea  \eq{step}  discussed in \Sec{DMdecay} afterward. 
Whether $T\gg m_\f$ is satisfied or not depends not only on the parameters  $y, m_\f, m_\psi$ but also on $z$. 
Thus $T\gg m_\f$ can be invalid today but valid in the early Universe, implying that there is parameter region $\D N_{\rm eff}[z_{\rm rec}] \simeq (0.01-1)$ in the early Universe while $E_F=m_\f/2$ today (right to the blue dashed line in Fig.\ref{fig:DMODS}).

\subsection{Predictions for the case without a chirality suppression}
\lac{sec:pred-no-chirality-suppression}
The 
slow but steady production of the dark radiation, 
a consequence of the self-annihilation of the condensate DM in the chirality-unsuppressed scenario,
 as well as the enhanced stability of the condensate,
may have interesting effects on cosmological evolution. 

\paragraph{Preliminaries for alleviating Hubble tensions}
\lac{Htension}
If $\D N_{\rm eff}[z_{\rm rec}]\ll 1$, then our DM scenario reproduces the ordinary $\L$CDM model predictions. 
Nevertheless, when $\D N_{\rm eff}[z_{\rm rec}]\sim \O(0.01-1)$ (corresponding to the purple shaded region in Fig.~\ref{fig:DMODS}), the model predicts 
a slight deviation from $\L$CDM which has the potential to alleviate the Hubble tensions.  
Let us qualitatively describe the effects that should help to alleviate the Hubble tensions. 

The analysis of the previous section and Fig.\ref{fig:DR} shows that our model realizes a novel decaying DM plus dark radiation scenario in which the dark radiation has self-interactions.
It has been discussed that cold DM decaying into dark radiation can partially alleviate the $H_0$ and $\s_8$-tensions~\cite{Aoyama:2014tga, Audren:2014bca, Enqvist:2015ara}. However, this scenario faces severe constraints arising from the late time changes to the DM density, which affect structure formation, enhance the late integrated Sachs Wolfe effect, and suppress the gravitational lensing effects~\cite{Bringmann:2018jpr,Enqvist:2019tsa,Simon:2022ftd, Alvi:2022aam}. These effects leave observable signatures in BAO, Sunyaev-Zel'dovich cluster counts, and CMB lensing.
In contrast, the DM condensate in our scenario has a suppressed ``decay" rate at late times, implying minimal impact 
on these observables such that those bounds should be
evaded (see Fig.\ref{fig:DR}). 

Furthermore, it is known that when $\D N_{\rm eff}[z_{\rm rec}]= \O(0.1\text{-}1)$  around the time of recombination, the $H_0, M_B$-tensions can be  relaxed~\cite{Mortsell:2018mfj,Riess:2019cxk,DiValentino:2020zio, Blinov:2020hmc}. This is because the extra radiation decreases the sound horizon size at the decoupling of CMB photons, which decreases the angular distance and thus alleviates the tensions. However, the amount of free-streaming dark radiation is severely constrained by the CMB data since it increases the Silk-damping and neutrino drag effects. On the other hand, if the dark radiation is self-interacting, it clusters more than the free-streaming radiation on small scales, counteracting the enhancement of the Silk damping effect.  As a consequence, the bound on the amount of dark radiation is relaxed if it is non-free-streaming~\cite{Blinov:2020hmc}.
The efficient self-interactions of the dark radiation in our scenario may inhibit free streaming.

Thus, for sizable couplings $y\sim {\cal O}(10^{-9})$, our scenario, which features a slow dark matter condensate evaporation along with substantial dark radiation with self interactions, contains the right elements to help to ease the Hubble tensions. 
A rigorous assessment of this possibility is quite non-trivial, requiring the use of a Boltzmann code to evolve the cosmological perturbations in order to make accurate predictions for cosmological observables, as well as a careful comparison of these predictions with various observational datasets, and goes beyond the scope of this work.

\paragraph{Suppression of small scale structure}

Another interesting prediction of our scenario is the absence of structures with large DM densities, whose formation is inhibited by DM evaporation. 
For instance, our scenario disfavors cuspy DM halo profiles for couplings that not too small.
One can naively estimate the maximal energy density of the DM by assuming that the dark radiation is homogeneous and has a temperature of $T[0]$. 
Then the maximal energy density persisting during the timescale $\D t$ satisfies 
\beq
\frac{y^4 m_\psi^2 \rho_\f^{\rm max}}{2\pi \max{[T[0], m_\f]}m^4_\f } \times \Delta t \sim 1.
\eeq
By taking $\Delta t = 1/H_0$ this gives  
\beq
\laq{rhomax}
 \frac{\rho_\f^{\rm max}}{\GEV{\rm cm}^{-3}}\sim \max{[0.2\(\frac{m_\f}{1\, {\rm meV}}\)^{8/5}\(\frac{ 10^{-9}}{y}\)^{16/5} \frac{m_\f^{8/5}}{m_\psi^{8/5}},1 \(\frac{m_\f}{1\,{\rm meV}}\)^{3}\(\frac{ 10^{-9}}{y}\)^{4} \frac{m_\f^2}{m_\psi^2}]}.
\eeq
where we have used \Eq{DRTt}. Note that $m_\psi/m_\f\ll 1/2$ gives an enhancement of the l.h.s. 
This naive estimation implies that the scenario may not lead to cuspy DM halo profiles, at least for large enough couplings.

Given  $\rho_{\f}^{\rm max}$,  we can estimate an upper bound on the DM mass in a finite sphere with a radius $r$, 
\beq
M^{\rm max}[r]=2.9\times 10^7M_\odot \(\frac{r}{300\rm pc}\)^3 \frac{\rho_{\f}^{\rm max} }{10 \GEV{\rm cm}^{-3}},
\eeq
where $M_\odot$ denotes the solar mass. 
For instance, 
a conventional dwarf spheroidal galaxy has a mass $M_{300}\sim 10^7 M_\odot$ within $r<300$ pc \cite{walker2009universal}.
This may set a bound \beq \rho^{\rm max}_{\f}\gtrsim (1\text{-}10)  \GEV {\rm cm}^{-3}.\laq{rhomax}\eeq 
While this is satisfied in most of the parameter regions, 
it may probe part of the region when the dark radiation is abundant. A more robust bound, beyond our preliminary estimate here, would require further detailed studies, including simulations of structure formation in our scenario with relatively large couplings. 
}
This could be particularly interesting given the anticipated future improvements in precision measurements of the masses of dwarf spheroidal galaxies as well as the (sub)halo mass function, e.g., using the Vera Rubin Observatory~\cite{LSSTDarkMatterGroup:2019mwo}.

\subsection{Parametric resonance and DM production}
\label{secpr}
So far, we have focused on the perturbative regime defined by \Eq{con1},
where we can use the Boltzmann equations to describe the system. 
 On the contrary,   
when \beq Q\gg 1,\eeq  $\psi$ acquires a large oscillating mass  $M_{\rm eff}(t)\simeq y \f{(t)}$,
the typical size of which is 
\beq
\laq{effmass}
\bar{M}_{\rm eff}\sim y \f_{\rm amp}= Q^{1/2} m_\f \gg m_\f,
\eeq
where $\f_{\rm amp}$ is the amplitude of the oscillating $\phi$ field. 
In this case, a perturbative description breaks down, and we cannot naively use the Boltzmann equation formalism. 
A similar system has been studied in the context of preheating after inflation~\cite{Kofman:1994rk,Kofman:1997yn, Greene:1998nh, Baacke:1998di, Greene:2000ew }. 
Here we follow the analytical discussion for preheating 
given in \cite{Kofman:1997yn} to quantify and describe the dynamics. 

One can consider the frequency of the fermion particle, $w_k$ of momentum $k$ at time $t$
\beq
w_k=\sqrt{k^2 +y ^2 \f[t]^2 }.
\eeq
The particle picture of the fermion can be justified when the adiabatic condition is satisfied 
\beq
\laq{cond}
\dot{w}_k\lesssim  w_k^2 .
\eeq
When this condition is violated with large $|\dot{\f}|$ , efficient $\psi$ particle production  occurs. 
Taking $\f=\f_{\rm amp}\cos[m_\f t],$ this violation can only occur at $\f\sim 0$, or more precisely $|\f|\lesssim (m_\f \f_{\rm amp}/y)^{1/2}$. The momentum
$k$ should also be small for \eq{cond} to be violated, i.e.,
\beq
\laq{fermi}
k\lesssim k_* \simeq \sqrt{y m_\f \f_{\rm amp}}= Q^{1/4} m_\f .
\eeq
As a result, a Fermi sphere with all modes less than $k_*$ is produced.
The Fermi momentum is thus
$
k_F\sim k_*$~\cite{Greene:1998nh,Greene:2000ew}.

Since $ k_F \ll \bar{M}_{\rm eff}$, the fermion is non-relativistic when the particle picture is justified with \eq{cond}. During most of the evolution} (and thus the time average of)
the energy density is given by
\beq
 \laq{para}
\rho_\p \sim \bar{M}_{\rm eff}n_{\p} = \frac{g_\p}{2\pi^2} Q^{5/4}m_\phi^4= \frac{g_\p}{2\pi^2} y^2 Q^{1/4}\rho_\f,
\eeq
where we used
$
n_\p\simeq \frac{g_\p}{2\pi^2}k_F^3. 
$
We get the scaling of the energy density of Fermi sea as $\rho_\p\propto a^{-15/4}.$

However, the simplification that $\f$ is described as a classical oscillating field should be treated with some care. 
For example, it is known that if there is a large quartic interaction of $\f$, ${\cal L}\supset -\lambda \f^4$, satisfying $y^2/\lambda\ll 1$, the system enters into a turbulence regime. This is found by performing a lattice simulation with a 2PI effective action in specific models~\cite{Berges:2010zv,Berges:2013oba}. However, this should not apply to our scenario if a small  
quartic coupling of typical radiative size 
$\lambda \sim y^4/16\pi^2$ is present.
The detailed numerical study of the $Q\gg 1$ regime for the quadratic potential is beyond our scope. 
In the following, we make use of the results of Refs.\,\cite{Greene:1998nh,Baacke:1998di,Greene:2000ew} with a quadratic potential of $\f$ in which back-reactions, such as the re-scattering effect, are neglected.

$Q\propto a^{-3}$ in the expanding Universe. Thus $Q$ may be larger than unity in the early Universe even if $Q\ll1$ at the later time. 
\begin{figure}[t!]
\begin{center}  
\includegraphics[width=145mm]{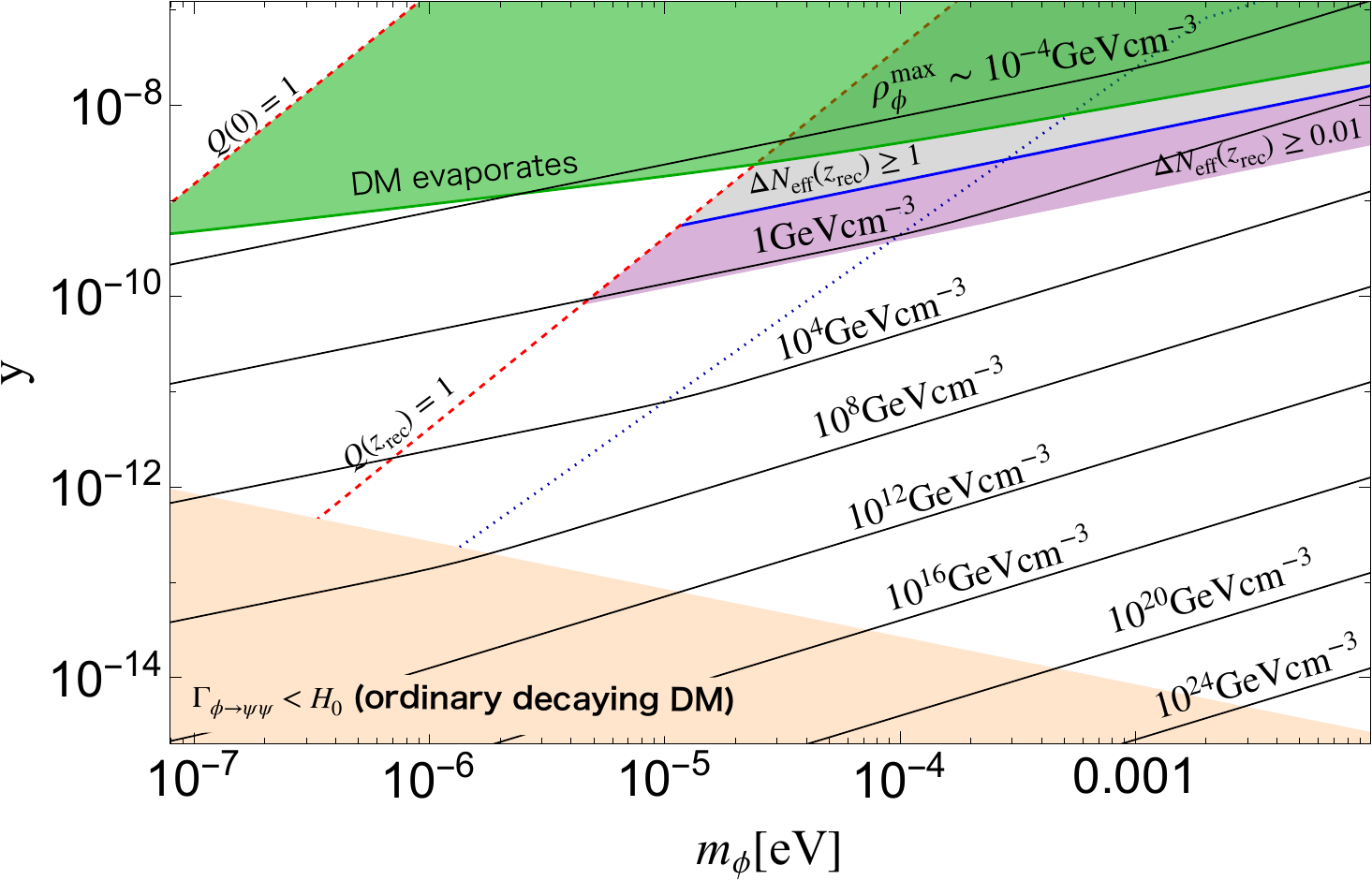}
\end{center}
\caption{Parameter region of DM stabilized by exotic fermions in $m_\f\text{-}y$ plane with $m_\f=5m_\psi$. Above the red dashed lines, broad parametric resonance is important at $z\leq z_{\rm rec}$ and $z=0$ from bottom to top, respectively. 
 Above the blue solid line, $\D N_{\rm eff}(z_{\rm rec})\geq 1$ and is disfavored.  
 In the purple region, 
 $\D N_{\rm eff}=0.01\text{-}1$ is predicted at $z=z_{\rm rec}$. 
 Below the blue dotted line, we get the Fermi sea with \Eq{step} in the present Universe.  Above it, the thermal component including the Fermi sea has a temperature $T[0]$ much larger than $m_\f/2$.  
The green region above the green solid line is disfavored since the comoving DM density may change significantly at $z<10^6$ (See \Eq{evae}). 
{We also show isocontours of the maximal local DM density in the present Universe, which is a specific prediction of our scenario, roughly estimated according to \Eq{rhomax}.}
  } 
\label{fig:DMODS}
\end{figure}
Due to the suppressed fermion energy density $\rho_\psi$ compared to that of DM $\rho_\f$ in the regime $Q\gg 1$, we can consider the conventional misalignment mechanism~\cite{Preskill:1982cy,Abbott:1982af,Dine:1982ah} for the production of the $\f$ abundance. That is, $\f$ starts to oscillate around the potential minimum when $m_\f \sim H$, and subsequently, the coherently oscillating scalar field forms a pressureless, nonrelatvistic fluid and thus serves as cold DM.  
The onset of oscillation is  at the photon temperature 
\beq 
\laq{osc}
T^{\rm osc}_\g \sim 300\GEV \(\frac{m_\f}{10^{-4}\EV}\)^{1/2} \(\frac{100}{g_\star}\)^{1/4},
\eeq 
from $H\sim m_\phi.$ From \Eq{para} and $\rho_\f=\rho_{\rm DM}^{\rm \L CDM}$, we get 
\beq 
\rho_\psi \sim 10^{-10}\(\frac{y}{10^{-9}}\)^{5/2} \(\frac{10^{-4}\EV}{m_\f}\)  \(\frac{T_\gamma}{300\GEV}\)^{3/4} \rho_\f,
\eeq 
and 
consequently $\rho_\psi < \rho_\f$ is satisfied with $T_\g <T_\g^{\rm osc}.$ 
Noting that $T_{\rm osc}\propto m_\f^{1/2},$ we always have $\rho_\p< \r_\f$ assuming the constraints \Eqs{massbound} and \eq{uppery} are satisfied. The fermion energy density
 $\rho_\psi$ is suppressed until $Q$ decreases to $\lesssim 1$ due to the expansion of the Universe, and following this we should use the expression of \eq{T}. 
This happens below 
\beq
T_\g^{\rm pert}\sim 5\EV  \(\frac{m_\f}{10^{-4}\EV} \)^{4/3}\(\frac{10^{-9}}{y}\)^{2/3}.
\eeq
Thus our previous discussion with \Eq{T} in the regime $z\lesssim T_{\g}^{\rm pert}/T_\g^{\rm today}-1$ with $T_{\g}^{\rm today}$ being the photon temperature today, 
  can be consistent with the misalignment mechanism.\footnote{
  It is conceivable that a full numerical study of the broad parametric resonance regime could reveal further nonperturbative effects which cause the DM condensate to evaporate. We emphasize that our scenario should still work in the perturbative regime $Q < 1$ together with a mechanism for lowering the initial redshift $z_{\rm ini}$ and dynamically delaying the onset of the $\phi$ oscillations (see, e.g., Refs.\,\cite{Daido:2017wwb, Takahashi:2019pqf, Marsh:2019bjr, Nakagawa:2020eeg}). In particular, a thermal potential may delay the onset of the $\phi$ oscillations, see, e.g.,~\cite{Batell:2021ofv,Batell:2022qvr,Murgui:2023kig}.
}

We must require that the DM does not evaporate too quickly after the 
end of the broad parametric resonance due to the perturbative interactions. 
To this end, we employ the condition 
\beq
\laq{evae} 
\rho_{\rm DS}[z_{Q=1}]< \rho_{\rm DM}[z_{Q=1}]
\eeq
with $z_{Q=1}$ defined by $Q[z_{Q=1}]=1$, i.e., $T_\g=T_{\g}^{\rm pert}$. Here $\rho_{\rm DS}$ is obtained using \Eq{T} evaluated in the $Q\ll1$ region. We 
impose
 this constraint if $z_{Q=1}<10^6$. 
If this condition is violated, too much of the DM energy density may be transferred into radiation 
during structure formation, cf.\cite{Sarkar:2014bca, Das:2020nwc}.\footnote{If $z_{Q=1}\gg 10^6$, it should be safe as long as the remnant at lower red shift is consistent with the DM abundance. Namely in the earlier time, the DM abundance decreases like in the WIMP case.}
This condition excludes the green region 
in Fig.\ref{fig:DMODS}. 
However, we emphasize that this constraint should be taken with caution.
When 
$Q\sim 1$ the 
system is non-perturbative, i.e., many $\leftrightarrow$ many processes may be as important as $1 \leftrightarrow 2, 2\leftrightarrow 2$ processes and there is also an oscillating effective mass of $\psi$,
which are not accounted for in our Boltzmann equation approach. These nonperturbative effects could in principle be studied using lattice methods.

\subsection{Cosmology and summary}

Let us summarize one of the possible cosmological histories for the scenario in which DM is stabilized by exotic BSM fermions. 
When $T_\g>T_\g^{\rm osc}$ (\Eq{osc}), $\f$ is frozen at some field value, which may be generated 
stochastically due to  inflationary quantum fluctuations. 
As the temperature drops, $\phi$ eventually starts to oscillate around its potential minimum when $T_\g \sim T_\g^{\rm osc}$, i.e., it is produced via the misalignment mechanism. 
 At this moment, $Q\gg 1$ and the system is in the regime of broad parametric resonance, resulting in the production of 
 a Fermi sea with energy density
 \eq{para}. Still, the energy density of the $\f$ condensate dominates over $\rho_\psi$, and $\f$ is stable~\cite{Greene:1998nh,Baacke:1998di,Greene:2000ew}.  As the universe expands, $Q \propto a^{-3}$ redshifts, and when $Q \sim 1$,  $k_*\sim m_\f$ and $\rho_\p\sim y^2 \rho_\phi\sim  m_\f^4$.
Afterward, the perturbative description 
accounting for the leading $1\leftrightarrow 2$ decay/inverse decay and $2\leftrightarrow 2$ scattering processes using  Boltzman equations is justified.
The results of the Boltzmann equation analysis can be found in Figs. \ref{fig:boltz} and \ref{fig:boltz2} with the initial condition of $T\sim m_\f$. The evolution of the system 
on timescales longer than $1/\G_{\rm kin}$ are relatively insensitive to the
initial conditions, and therefore, 
the imprints of the initial broad parametric resonance 
phase are largely erased. 

The parameter space for DM stabilized by exotic fermions in the $m_\f-y$ plane is displayed in Fig.\,\ref{fig:DMODS}, where we take $m_\psi/m_\f=1/5.$
In the green region 
the DM density changes by an $\O(1)$ amount soon after the broad parametric resonance ends, which takes place at $z<10^{6}$, and may thus be excluded.  
The gray region above the blue solid line is disfavored since $\D N_{\rm eff}(z_{\rm rec})>1.$ 
In the 
purple region we predict 
slowly produced interacting dark radiation (see Fig.\,\ref{fig:DR}) with $\D N_{\rm eff}(z_{\rm rec})=0.01-1$, which
may 
potentially alleviate the Hubble tensions, suppress the formation of DM structures with large densities, 
and can be probed 
in the future
CMB and BAO experiments~\cite{Kogut:2011xw, Abazajian:2016yjj, Baumann:2017lmt}.   
Above the lower red dashed line the $Q$ parameter at the recombination era, $z\sim z_{\rm rec}$, is greater than unity and the dark radiation production is suppressed.  
Above the upper red dashed line the system is in the broad parametric resonance regime  in  
the present Universe.

Above the blue dotted lines,  2-to-2 scattering processes are important and $T \gg m_\f/2$ in the present Universe. 
Below the blue dotted line the naive discussion in \Sec{DMdecay}, which accounts only for 1-to-2 decay/inverse decays processes (including Pauli blocking) but neglects scattering, applies in the present Universe (but not necessarily the earlier Universe) 
since the 
$\f_c$ annihilation is not efficient. In this case, the system today has a Fermi sea  component, \Eq{step}, due to the Pauli blocked decay of $\f$.

Finally, in the orange region, the DM has a longer lifetime than the age of the Universe in vacuum, and thus corresponds to the standard decaying DM scenario.

\section{DM stabilized by C$\nu$B}
\lac{DMONS}
 So far, we have investigated DM stabilization due to Pauli-blocking in a minimal model containing an exotic light BSM fermion.
It is natural to consider the possibility that the light fermion is instead a neutrino,
either a left-handed neutrino or a right-handed neutrino, i.e., $\psi=\nu_L$ or $\nu_R$.
The $\psi$ in the previous section can be straightforwardly identified with $\nu_R$ without changing our previous conclusions 
provided $\n_R$ has a suppressed mass and not too large mixing with the ordinary neutrinos. 
The more interesting case is $\psi$ as $\nu_L$, which we consider in the following.

The interaction Lagrangian is given as~\footnote{Neutrino couplings to scalar dark matter have been investigated in various other contexts, see, e.g., Refs.~\cite{Berlin:2016woy,Krnjaic:2017zlz,Brdar:2017kbt,Capozzi:2018bps,Huang:2018cwo,Farzan:2018pnk,Dev:2020kgz,Losada:2021bxx,Chun:2021ief,Dev:2022bae,Gherghetta:2023myo,ChoeJo:2023ffp}. 
} 
\beq 
\laq{lagnu}
{\cal L}\supset -\frac{1}{2} y_{ij} \f (\bar{\n}^c_L)_i  (\n_L)_j+{\rm h.c.}, 
\eeq
where $i={1,2,3}$ is the index in the mass eigenbasis, e.g., in the normal ordering case, the lightest, heavier, and heaviest neutrinos, respectively. 
We assume the normal hierarchy, i.e., normal ordering with a small lightest neutrino mass, 
\beq 
m_{\nu_1}< m_\f/2,
\eeq 
so that DM is stabilized as a consequence of the Pauli-blocking of its decays to the lightest neutrino.  
We also assume a generic flavor structure, $y_{ij}\sim y$, and as such we omit the flavor index in the following. 

There are several 
aspects of this scenario that are different from the minimal model discussed in \Sec{DMODS}. 
First, there is the presence of the C$\nu$B from the conventional Big Bang cosmology that contributes 
$ N_{\rm eff}\sim 3$, which was confirmed in the 
CMB and BAO observations~\cite{Planck:2018vyg}.
Since the C$\nu$B at early times should not deviate significantly 
from the standard cosmological predictions, 
 in the following, we approximate that 
$T_\nu$ of the plasma is the usual neutrino temperature, $T^{\rm SM}_{\nu}$, i.e., $T_\nu[z>z_{\rm rec}] \sim T_{\nu}^{\rm SM}[z>z_{\rm rec}]$, before the CMB era in most of the discussion. 
A slight modification to the standard cosmology will be also discussed. 
On the other hand, in the later epochs the C$\nu$B 
may be altered significantly without violating the CMB and BAO constraints. 
As we will see, the change in the late time C$\nu$B spectrum is a natural prediction of our scenario.

Second, the extra interaction of the C$\nu$B with $\phi$ may inhibit the free streaming of neutrinos, in contradiction with inferences from CMB and BAO data~\cite{Blinov:2020hmc}. 
This leads to 
a new constraint on the coupling $y$. 

Lastly, 
the heavier neutrinos are expected to decay during the age of the Universe, even if the coupling of $y$ is not very large. 
This outcome is natural since we focus on the parameter region where the $\f$ proper lifetime (in vacuum) is shorter than the age of the Universe, and given the assumption of a generic flavor structure, a coupling of the same order leads to an even faster decay of the heavier neutrinos ($\nu_3\OR\nu_2$).
Eventually, the $\f_c$ decay is Pauli-blocked in our universe, 
but the decay of $\nu_{3,2}$ is not. 

We  discuss these three points in more detail in the following.

\subsection{C$\nu$B free-streaming and DM stabilized by neutrino sea }
\lac{CnuB-free-stream}
Under our assumption of 
a generic flavor-violating structure in the $\phi$-neutrino couplings, $y_{ij}\sim y$,  
a heavier neutrino in the rest frame can decay into $\f$ and a lighter neutrino with a decay width  
\beq
\tau_{\nu_{2,3}}^{-1}\equiv \G_{\nu_{2,3}}\sim\frac{y^2 m_{{\nu}_{2,3}} }{16\pi} 
\eeq
where $m_{\nu_i}$ is the neutrino mass of $(\nu_L)_i$, and we neglect the mass of the decay products. 
When the lifetime is shorter than the age of the Universe, the heavier neutrinos are absent in the present C$\nu$B, which then consists of only the lightest neutrino.

Such decays 
may be constrained as they can hinder 
neutrino free-streaming via the process $\nu_3 \leftrightarrow \f + \nu_{1,2}.$ 
These reactions 
should not be too frequent around the recombination era.  
To study  free-streaming it is important to check the evolution of the anisotropic component of the plasma. 
For the decay of a relativistic particle of energy $E_{\n_3}$ and mass $m_{\nu_3}$ in the cosmic frame, daughter particles have  momenta in the collinear angle $m_{\nu_3}/E_{\n_3}$. An inverse decay by the annihilation of a particle with the ambient plasma produces heavier particles also in a collinear angle  $\sim m_{\nu_3}/E_{\n_3}$. Thus the initial momentum diffuses to the whole momentum space after order $(m_{\nu_3}/E_{\n_3})^{-2}$ 
reactions, similar to a random walk. Thus, it has been argued that the timescale for momentum transport
is further suppressed by the diffusion factor together with the boost factor $(m_{\nu_3}/T)^2\times m_{\nu_3}/T$~\cite{Hannestad:2005ex,Archidiacono:2013dua,Basboll:2008fx,Escudero:2019gfk}.
Moreover, recently it was found that there is an additional suppression numerically (and understood analytically under several assumptions) by $(m_{\nu_3}/T)^2$ due to a cancellation of the collision terms of the relevant order in a derived evolution equation for the total anisotropic stress~\cite{Barenboim:2020vrr}. 
Here we adopt the bound~\cite{Barenboim:2020vrr} (see also \cite{Hannestad:2005ex,Archidiacono:2013dua,Basboll:2008fx,Escudero:2019gfk})\footnote{This is the case with a generic flavor violation for the $\phi$ coupling. In the case of the Majoron or $B-L$ Higgs boson, the neutrino decay of $\nu_{2,3}$ is suppressed. 
But there is a free-steaming constraint coming from the reaction $\phi \leftrightarrow \nu_1 \nu_1$.
The bound can be similarly obtained~\cite{Barenboim:2020vrr} by using the timescale $\Gamma^{-1}_{\f \to \nu_1 \nu_1} \(\frac{m_\f}{T}\)^{-5}$.}
\beq
\laq{non-free}
y \lesssim 10^{-9} \(\frac{m_{\nu_3}}{0.05\EV}\)^{-3}.
\eeq
If this bound is satisfied, then 
two-to-two processes involving neutrinos and thermalized $\phi$ particles 
will also be suppressed~\cite{Barenboim:2020vrr}. 

Let us argue that in our scenario, the condition \eq{non-free} is sufficient to ensure neutrino free-streaming, including also the effects of the specific scattering reactions involving the DM condensate $\f \f_c \leftrightarrow \nu_1\nu_1$ and $\f_c \nu_1\leftrightarrow \f\nu_1$. 
The reaction rate \eq{scat} is not the appropriate one for damping anisotropic stress since the outgoing particles are collinear. By including the diffusion factor $(E_{\rm cm}/T)^2 \sim m_\f/T$, 
we estimate the 
transport rate to be $\G_T \sim \frac{m_\f}{T} \G_{\rm scat} \left(\frac{m_\f}{T}\right)^{n-1}$, where $n-1 \geq 0$ is a number that represents the possible further cancellation in the evolution equation of the total anisotropic stress including the potential Pauli blocking effect. We find that $\G_T \sim y^4 \frac{T^2}{m_\f} \left(\frac{m_\f}{T}\right)^{n-1} \frac{\hat n_\f m_\f}{T^4}$. 
Since $\hat n_\f m_\f \sim T^4$  around matter-radiation equality and thus recombination, and $n > 0$, the transport rate is always smaller than the Hubble rate in this parameter region.

It should be noted that if \Eq{non-free} is satisfied, the usual interaction rate $\frac{y^2}{4\pi} m_{\nu_3}^2/T_{\nu}$ can only become faster than the Hubble rate much after the BBN era. Therefore, the entropy of the C$\nu$B shortly after decoupling around the BBN epoch is the same as that of the SM case.

\subsection{Spectrum and composition of C${\nu}$B}
Our scenario predicts 
a non-trivial spectrum and composition of the C$\nu$B as a consequence of neutrino decays and  neutrino production from the $\f$ condensate after 
recombination. 
The spectrum of the C$\nu$B is, in principle, measurable 
at future C$\nu$B detection experiments, and thus provides a novel probe of our model.
In the following, we will consider two scenarios.
In the first part (\Sec{1}), we consider the case in which the DM 
annihilation is
chirality suppressed.
In the second part (\Sec{2}), we consider the chirality unsuppressed case. 
The second case is 
similar in most respects 
to the first case, but when the coupling is large we may potentially alleviate the Hubble tension. 
As we will see, both scenarios typically feature 
non-standard components of the C$\nu$B which may be probed in the future.

\subsubsection{Case of chirality suppression 
}
\lac{1}
In the chirality suppressed case, $m_{\nu_1}/m_\f\ll 1$, 
the lifetime of $\phi$ can be estimated from \Eq{groT} to be
$
\sim (y^4 T)^{-1}\sim 10^{7}\times 13.8 \, {\rm Gyr}\(\frac{y}{10^{-9}} \)^{-4} \frac{T^{\rm SM}_{\nu }[0]}{T}.
$

There are two sources of the C$\nu$B in the present Universe. One is the standard primordial source of neutrinos which originates from the hot Big Bang. We stress that in our scenario the spectrum, however, 
 may be quite different from the standard one because the additional interaction involving $\phi$ leads to the decays of neutrinos. The effect also depends on the size of the coupling $y$, thus we will examine two regimes for its strength in the following.
 The other source of the C$\nu$B is the $\phi$ condensate decays, which will be discussed lastly, and this contribution should be added to the primoridal one.

\paragraph{C$\nu$B component from primordial neutrinos at weak coupling}
If $
y\lesssim 10^{-14}
$ the heaviest neutrino decay happens at $T_\nu\lesssim m_{\nu_3}\approx 0.05\EV.$  Thus $\nu_3$ becomes non-relativistic first and then decays via
\beq\laq{3decay}\nu_3\to \f({\rm boosted}) +\nu_{1,2}.\eeq
In this case, the decays happen out of equilibrium.
Interestingly, since the Pauli blocking effect can be neglected for the boosted $\f$ decay,\footnote{In a small portion of the parameter space, there is also the possibility that the further decay of $\nu_2,\f(\rm boosted)$ does not happen within the age of the Universe due to the time dilation effect, which we do not account for in our order-of-magnitude estimate.} 
\beq 
\laq{2decay}\nu_2 \to \f(\rm boosted)+\nu_1, ~~\f(\rm boosted)\to 2\nu_1
\eeq
can happen subsequently. Here $\nu_2$ may be either the product of a $\nu_3$ decay or a relic that was produced around BBN. 
Assuming $\nu_3$ decays to $\nu_2\phi$ and $\nu_1\phi$ with the same rate, the total comoving number density of the neutrinos increases from $(n_{\nu,1}+n_{\nu,2}+n_{\nu,3})a^3\to (8n_{\nu,1})a^3$. 
Some event rates for the neutrino interaction with electrons are proportional to $|(U_{\rm PMNS})_{e i}|^2=\{0.7, 0.3, 0.02\}$, with $U_{\rm PMNS}$ being the PMNS matrix.  In particular, this is the case for the tritium capture rate in the PTOLEMY project~\cite{Betti:2019ouf, McKeen:2018xyz,Long:2014zva}. 
When the heavier neutrinos all decay into the lightest one, the event rate can be
enhanced by a factor\footnote{ 
In contrast, for the inverted hierarchy case with a negligible lightest neutrino mass, the event rate is suppressed by $0.02\times 8/(0.7+0.3+0.02)\sim 10\%$.  
} \beq \frac{0.7\times (1+3+0.5(3+5))}{(0.7+0.3+0.02)}\sim 6 ~~~~~~ ({\rm typical~} E_\nu= \O(0.01)\EV)
\laq{det1}
\eeq 
compared to the SM case. 
Here we used the fact that the event rate does not depend much on the energy of the neutrino. 
{Here and hereafter, the bracket after the formula for the enhancement of the C$\nu$B number density denotes the typical energy of the component.}
Thus, the precise measurement of the C$\nu$B may provide a probe of our scenario. 
The expected energy resolution of the PTOLEMY experiment is $0.01-0.1\EV$~\cite{Betti:2019ouf}, which favors energetic neutrinos. 
In our case, the part of the numerator $3+0.5(3+5)$ represents the neutrinos produced in 
the cascade processes of \Eqs{3decay} and \eq{2decay}, which are boosted in comparision to the primordial component. These have energies as large as $\sim (1/4-1/2) m_{\nu_{2,3}}=\O(0.01)\EV$  if $\n_2,\n_3$ decays around the present Universe such that 
we can neglect the effects of redshift.

\paragraph{C$\nu$B component from primordial neutrinos at moderate coupling} 
When 
$ y\gtrsim 10^{-14}$
the decay and inverse decay happen thermally at $T_{\nu}\gtrsim m_{\nu_3}$.
All neutrinos 
reach 
quasi-thermal equilibrium with $\f$ well
after BBN. 
The comoving entropy stored in the neutrino-$\phi$ system is conserved and  
is equal to the SM value carried by the three neutrinos following their decoupling around the BBN era at which point the $\phi$ interaction is frozen. 
Later, after the decay/decoupling of all the heavier neutrinos, we have the relation 
\beq 
(1+ 7/4) \frac{2\pi^2}{45}(T_{\nu}[0])^3\simeq 3\times(7/4) \frac{2\pi^2}{45}(T_{\nu}^{\rm SM}[0])^3\eeq 
from comoving entropy conservation. Here we have assumed that the mass of $\phi$ is smaller than the temperature. We will discuss the case of heavier $\phi$ later. 
As a result, the final comoving neutrino number density 
of the stable lightest $\nu_1$ increases by  
\beq
\(\frac{T_{\nu}[0]}{T^{\rm SM}_{\nu}[0]}\)^3 \simeq \frac{21}{11}, 
\eeq
and thus the detection rate is enhanced by a factor of 
\beq\frac{ 0.7 \times (21/11)}{(0.7+0.3+0.02)}\sim 1.3 
 ~~~~~~
 ({{\rm typical~} E_\nu= \O(T^{\rm SM}_\nu[0])}).\laq{det2}\eeq

In addition, for $10^{-9}\gtrsim y\gtrsim 10^{-13},$ the thermalization of $\f-\nu$ system happens before recombination, meaning that the energy density of the neutrino sector is slightly different 
from $\Lambda$CDM. 
Thus, it may be probed via measurements of $\Delta N_{\rm eff}$.
To perform the estimate we can again make use of the comoving entropy density conservation. 
Near the recombination era, all the three neutrinos as well as $\phi$ are relativistic, implying that the 
 relativistic entropy degrees of freedom in this sector are  $g_{s\nu}\simeq 3\times 2\times\frac{7}{8} +1$ compared to the SM case $3\times 2\times\frac{7}{8}$. 
 This leads to 
\beq 
(1+ 3\times 7/4)(T_{\nu}[z_{\rm rec}])^3 \frac{2\pi^2}{45}\simeq ( 3\times 7/4)(T^{\rm SM}_\nu[z_{\rm rec}])^3 \frac{2\pi^2}{45}.
\eeq
We obtain \beq \laq{negaNeff} \Delta N_{\rm eff}[z_{\rm rec}]\approx -0.17.\eeq
A deviation of this size 
can be tested in future CMB and BAO experiments~\cite{Kogut:2011xw, Abazajian:2016yjj, Baumann:2017lmt}. We note that the decrease in $\Delta N_{\rm eff}$ during the CMB era does not change the prediction \eq{det2}, consistent with 
comoving entropy conservation. 

\paragraph{C$\nu$B components from DM decay}

Since the DM stability in the present Universe is a consequence of the Pauli-blocked decay to the lightest neutrinos, and, at relatively small coupling strength the $\phi$ annihilation is negligible, we have an extra component of the C$\nu$B corresponding to the Fermi sea produced by the decay of the condensate. This component is characterized by the Fermi momentum of $m_\f /2$ and number density $n_{\nu_1} \sim (m_\f/2)^3/6\pi^2$, and therefore it dominates over the other components when $m_\f \gg T_{\n}^{\rm SM}[0]$.
In this case, the interaction rate of an electron with C$\nu$B 
is enhanced by a factor 
\beq
\laq{enhance}
\sim 10^3\( \frac{m_\f}{5\,\rm meV}\)^3 ~~~~~~~({\rm typical~} E_\nu = \O(m_\f)),
\eeq 
which is the ratio of the present-day neutrino number density to the standard one. 
Although the energy of these neutrinos is relatively small compared to the component from the heavy neutrino decays due to the bound  \Eq{massbound}, there can be several orders of magnitudes enhancement
in the event rate in the future C$\nu$B experiment. 
Detecting this component is the smoking gun signal of our scenario.

\subsubsection{Case without chirality suppression at {large coupling}}
\lac{2}
For the scenario in which  $\phi_c$ self-annihilation is not chirality suppressed, differences from \Sec{1} only appear for relatively large Yukawa couplings, $y\sim 10^{-9}$. 
In this case, the energy density of the C$\nu$B may be slightly enhanced around the recombination era due to the slow $\f_c$ evaporation. 
Following the discussion in \Sec{massive}, the C$\nu$B temperature is close to the temperature \eq{T} at recombination.  
At later times, the temperature follows the attractor 
solution, e.g., at $z=0$ we have \Eq{Neff}. 
Thus when the $\nu_1$ radiation from $\phi_c$ evaporation is produced in comparable amounts to the primoridal C$\nu$B, the evolution of the neutrino radiation follows the discussion in \Sec{massive}, resulting in a slight enhancement of the total C$\nu$B at $z=0$, provided the DM is light enough. 
{In particular, near recombination we can obtain $N_{\rm eff}[z_{\rm rec}]\sim 3-4$ while today, analogously} to \Eq{Neff}, we get $N_{\rm eff}[0]\sim 10 N_{\rm eff}[z_{\rm rec}].$
Thus we expect an enhancement of the C$\nu$B detection rate by as much as a factor of 
\beq\laq{enhance2}
\sim \frac{0.7\times (4\times 10)^{3/4}}{0.7+0.3+0.02}\sim 10 ~~~~~~~({\rm typical~} E_\nu= \O(T_\nu^{\rm SM}[0]) ).
\eeq
Such an enhancement of this kind occurs in the narrow blue-shaded region in Fig.~\ref{fig:DMONS}. However, when $m_\f>10^{-4}\EV$, the degenerate neutrino sea from the direct decay of $\f_c$ is more important, as in the chirality suppressed case.

For $y\ll 10^{-9}$, the additional neutrino radiation from $\f_c$ evaporation is negligible. 
Still, along with the primordial component, we predict the novel C$\nu$B components originating from neutrino decay as well as DM decay, 
similarly to the chirality-suppressed case discussed in the previous subsection.

Similarly to \Sec{sec:pred-no-chirality-suppression}, there is an interesting possibility that our scenario, with relatively sizable couplings $y\sim 10^{-9}$ (see Fig.~\ref{fig:DMODS}), may help to alleviate 
the Hubble tension. 
In this case, additional 
neutrino radiation is produced near recombination from $\f_c$ evaporation. 
The produced radiation with the standard neutrinos has a shorter free-streaming length than usual. It was pointed out that a majoron (scalar) coupled to neutrinos may relax the Hubble tension by introducing extra neutrino radiation $\D N_{\rm eff}[z_{\rm rec}]\sim 0.5$~\cite{Escudero:2019gvw, Escudero:2021rfi}. In our scenario, extra radiation corresponding to $\D N_{\rm eff}[z_{\rm rec}]\sim 0.5+0.17\sim 0.7$ can be provided naturally by $\f_c$ evaporation, where the additional amount $~0.17$ is needed to compensate the negative value of \Eq{negaNeff} caused by the late thermalization of $\f$ particles. 

In summary, in the case that the DM condensate annihilation to neutrinos is not chirality suppressed, we have a special parameter region (corresponding to the narrow blue shaded range in Fig.~\ref{fig:DMONS})
 where the $\phi_c$ evaporation produces extra neutrino radiation, which may potentially alleviate the Hubble tension and be 
probed in C$\nu$B detection experiments. When the DM mass is larger than about $0.1\,$meV, the Hubble tension may still potentially be addressed by this additional neutrino radiation, 
however this component is less important in the context of the C$\nu$B detection compared to the degenerate 
neutrino sea originating from the direct decay of DM in the present Universe. 
In both cases, we also predict an extra C$\nu$B component from the decays of the heavier neutrinos.
Below the blue region, we have the essentially same predictions as in the chirality suppressed case.

\subsubsection{Discussions on perturbativity}
The parameter space of DM stabilized by the C$\n$B in the $m_\f-y$ plane is 
presented in Fig.\,\ref{fig:DMONS}. We take $m_{\nu_1}/m_\f=1/5.$
Above the red solid line, the neutrino is massive during the BBN epoch due to the large $\phi$ amplitude, which will be discussed at the end of this section.
We note that the boundaries of these regions may vary to some extent 
depending on the component values of the coupling matrix
 Above the red dotted line, the $Q[z_{\rm rec}]$ parameter at the recombination becomes larger than unity and the Boltzmann equation at this era becomes non-perturbative. 
 Analyzing this regime may require a different approach such as e.g., lattice methods.

\begin{figure}[t!]
\begin{center}  
\includegraphics[width=135mm]{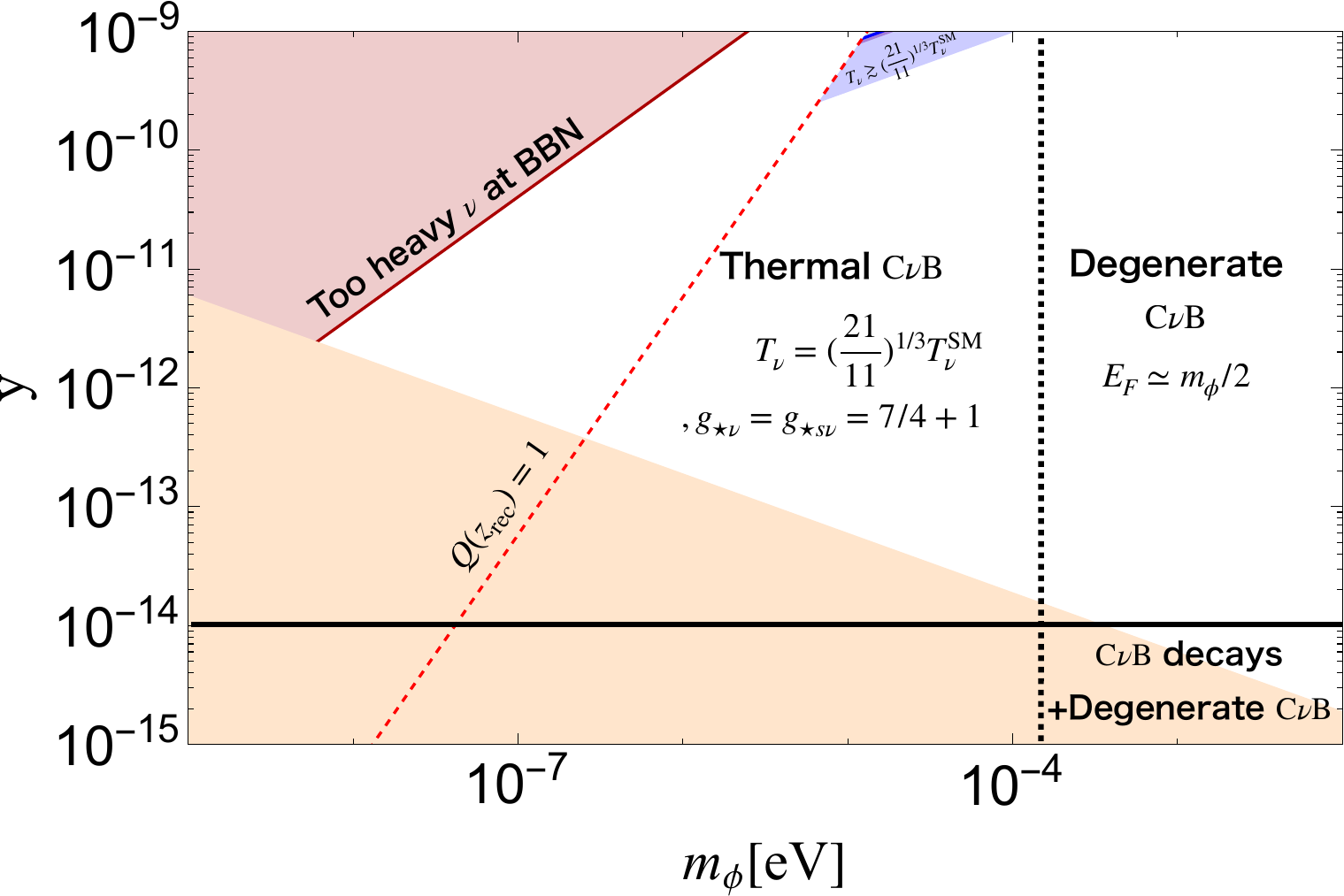}
\end{center}
\caption{The prediction for the present universe C$\nu$B with the DM stabilized by C$\n$B in $m_\f-y$ plane. We take $m_{\nu_1}=1/5 m_\phi.$ 
In the top blue-shaded region, the extra radiation from the DM evaporation increases the C$\nu$B-electron interaction rate by at most $\sim 10$ (see the discussion around \Eq{enhance2}). 
Below it, the $\rm C\nu B$ is thermalized, but we do not have the extra radiation but an enhancement of $\sim 1.3$ (see the discussion around \Eq{det2}). In both cases, the C$\nu$B is composed of a plasma of $\nu_1$ and $\f$ with the chemical potential of $\mu_1=m_\f/2.$ In the right top, the C$\nu$B is composed of the ``degenerate" $\nu_1$ from the DM decay. We can have several orders of magnitude of enhancement of the detection rate (see the discussion around \Eq{enhance}). In the bottom right region, the C$\nu$B is composed of boosted $\nu_1$ from $\nu_3,\nu_2$ and boosted $\f$ decays. The enhancement can be at most a factor of $6$.  
Above the red region above the red line, the averaged neutrino masses become heavier than the temperature at the BBN.
The other colored regions are the same as Fig.~\ref{fig:DMODS}.
  } 
  \label{fig:DMONS}
  \end{figure}

\subsection{Production mechanism of $\f$ stabilized by C$\nu$B.}

Before concluding this section, we discuss a constraint that is relevant if the DM condensate is produced through the misalignment mechanism. 
In this scenario, the onset of the oscillations takes place at the temperature given by \Eq{osc}. 
This is much higher than the BBN temperature and may even be higher than the electroweak phase transition temperature. 
The effective neutrino mass $\bar{M}_{\rm eff}$ during these early times is given by \Eq{effmass}.
In particular, one should avoid non-relativistic neutrinos during the CMB and BBN epochs, leading to the condition
$\bar{M}_{\rm eff}\lesssim T_\g$ at $T_\g=1\MEV$. This translates to the bound
\beq
\laq{yBBN}
|y|\lesssim 4\times 10^{-9}\(\frac{m_\f}{10^{-5}\EV} \).
\eeq

Furthermore, we note that at such early times and high temperatures that the free energy density of neutrinos sources a finite temperature contribution to the $\phi$ potential. Generically, the thermal mass of $\phi$ in this early epoch will dominate over the bare mass, such that $\phi$ no longer redshifts as matter during this era. However, once the temperature drops sufficiently so that the bare scalar mass again dominates, $\phi$ will behave as matter. It is also possible to imagine that the scalar potential is suitably modified so that $\phi$ oscillations do not begin until some later epoch following BBN. 
In this sense, the bound \eq{yBBN} is model-dependent.

 For energies and temperatures larger than the weak scale, we must find a reasonable UV origin of the Lagrangian \eq{lagnu}, consistent with the gauge symmetries of the SM.
The UV model should be consistent with neutrino mass 
generation, 
and furthermore, the thermalization of $\f$ via the interactions among $\f$, neutrinos (or left-handed lepton), and gauge bosons must be avoided. 
 Consistent renormalizable models utilizing
 the misalignment mechanism can be constructed by introducing right-handed neutrinos, and we present one such example model 
 in Appendix~\ref{sec:ap1}). 
An alternative possibility to evade these problems is to consider other mechanisms for the production of the DM $\f$, such as inflaton decays~\cite{Moroi:2020has,Moroi:2020bkq}.

\section{Generalizations to other models and signatures}
\lac{gene}
Here we discuss some possible generalizations to other models that may realize the mechanism of DM stability from Pauli blocking, as well as some additional potential signatures that may be present depending on how the dark sector couples to the SM degrees of freedom. 

\paragraph{Other models with DM stability from Pauli blocking} 
In the minimal model presented in \Sec{DMODS} 
we have assumed that $\psi$ is stable. However, this is mainly for simplicity. Our scenario should work equally well even if $\psi$ decays to lighter states. This is because  
the $\psi$ decay products must include at least one fermion, and thus the decay of $\psi$ will be Pauli-blocked as well.

Our mechanism for DM stabilization can also be applied to more generic interactions than the ones considered in this work. 
For instance, instead of \Eq{int} one can consider the interaction $\f \bar{\psi} \Psi$, where $\psi$ is a light or massless fermion 
while $\Psi$ is a heavy fermion with mass $M_\Psi$. The DM $\phi$ decays, filling the $\psi$ and $\Psi$ Fermi spheres such that the decay is eventually Pauli blocked. Interestingly, the corresponding Fermi momenta can be made arbitrarily small by taking the mass splitting $m_\f-M_\Psi$ to be small. In this limit,  the energy transferred from the DM to the fermions $\psi$ and $\Psi$ also becomes small. 
In comparision to the minimal model of \Sec{DMODS}, the advantage of this scenario is that one can have a relatively heavy $\f$ at the expense of a tuning between $m_\f$ and $M_\Psi.$\footnote{Such tuning can be related with the radiatively induced neutrino mass in certain UV models~\cite{Boehm:2006mi, Ma:2006km, Yin:2017wxm}. }
 In this case, however, we must require that the couplings of $\Psi$ to other lighter particles are small so that $\f$ decays via $\Psi$-mediated interactions are  ineffective. 

Instead of a primary two body decay that stabilizes DM, we may 
have three body decay channels involving fermionic decay products. These decays can also become Pauli blocked and DM can be stabilized. 
As an interesting example, 
consider the Lagrangian $\f \bar{\psi} \sigma^{\mu \nu}\psi F_{\mu \nu}$, where $F_{\mu \nu}$ is the photon field-strength.
In this scenario, the DM continuously produces photons as it replenishes the Fermi sea during the expansion of the Universe. 
On the other hand, a boosted $\phi$ (e.g., a dark radiation component) can decay into the fermions and a photon with substantially higher energies. The two distinct sources of photons 
can thus potentially provide an interesting signature of the mechanism. The coupling to photons may lead to further constraints from early universe cosmology, which should studied carefully.

In this work we have considered a scalar DM candidate, but our mechanism should also work for the case of vector boson DM coupled to fermions. 
Most discussions do not change qualitatively. 
A difference is that the annihilation of the DM is not chirality suppressed even in the massless limit of the decay product.

\paragraph{Indirect signatures from other couplings}
So far in this scenario, we have neglected the possible couplings of $\phi, \psi$ to SM particles, but they can be present on general grounds and may lead to additional indirect signatures of the dark sector. Here we mention two possibilities for illustration. 

The symmetries of the theory, \Eq{int}, allow higher dimensional interactions of the form $\frac{1}{M}H L \phi \psi \supset \frac{v }{M} \nu \phi \psi,$ with $H,v, \AND L$ being the SM Higgs doublet, its vacuum expectation value, and a lepton doublet, respectively.
Such couplings cause the C$\nu$B to decay to the lighter dark sector particles. Interestingly, the lifetime of the heaviest SM neutrino can be below the age of the Universe when the scale $M\lesssim M_{\rm pl}$.\footnote{More generally, this fact shows that C$\nu$B search may provide a nice opportunity for testing a light dark sector with a fermion and a scalar, even if direct couplings are only generated by gravitational effects. 
For instance, a lighter dark SM  
could be probed if the ordinary neutrino decays to a dark neutrino and dark Higgs boson.} Precisely measuring the C$\nu$B and finding the absence of the heaviest neutrino may probe this scenario.

Another interesting possibility arises if $\phi$ 
couples to a pair of photons, albeit with a decay rate that is highly suppressed. This scenario is natural if the $\phi$-photon coupling arises from a loop effect, such that it is highly suppressed compared to the $\phi$-fermion (\Sec{DMODS}) or $\phi$-neutrino coupling (\Sec{DMONS}). 
In the case of 
DM stabilized by the C$\nu$B, we have two sources of photon signals in the present Universe. One is from the non-relativistic DM rare decay to photons, with photon energy of half the DM mass (neglecting the redshift). The other is from the  decay of $\phi$ produced from the decay of a neutrino. 
For the second source, the resulting typical energy of photon is around  half of the mass difference of the neutrinos between the initial and final states. 
Interestingly, the latter energy does not depend much on the mass of $\phi$ if it is small enough. 
This process can be probed by C$\nu$B experiments searching for neutrino radiative decay such as COBAND~\cite{Takeuchi:2021atx}.

\section{Conclusions and outlook}

In this work we have explored the novel implications of the hypothesis that DM stability is a consequence of the Pauli exclusion principle. 
We studied a simple model containing a sub-eV scalar $\phi$ DM candidate with a Yukawa coupling to  a lighter exotic fermion $\psi$.  
Even though the scalar DM proper lifetime in vacuum, $1/\Gamma(\phi \rightarrow \psi \psi)$, is much shorter than the age of the Universe, the available fermionic $\psi$ states produced in the decay are rapidly filled and the decay is Pauli blocked. We have considered not only the impact of decay/inverse decay processes on the evolution of the scalar DM condensate, but also the relevance of scattering processes and parametric resonance effects, finding a substantial parameter space in which the effective lifetime of the DM condensate is much longer than the age of the Universe. In particular, we have demonstrated that for relatively large couplings, scattering processes can be rapid enough to produce a substantial quasi-thermal dark sector component behaving as (self-interacting) dark radiation, which can potentially help to relax the Hubble tensions. Future precision CMB and BAO observations will provide additional tests of the scenario.  
Furthermore, a cutoff in the densities of DM structures is expected, motivating future work on simulations of structure formation, which may help furnish an additional robust probe of the scenario. 

Perhaps the most natural candidate for the light fermion is the SM left-handed neutrinos. In addition to the predictions of the exotic fermion scenario summarized above, this scenario predicts significant modifications to the C$\nu$B at late times, both in terms of its composition and energy spectrum. There are several possible effects that may distort the C$\nu$B, including the cascade decays of the heavier SM neutrinos as well as DM evaporation and/or decay. In particular, in a large fraction of the viable parameter space, the interaction rate of the C$\nu$B with an electron is significantly enhanced. This opens up the possibility of testing our scenario through precise measurements of the C$\nu$B with experiments such as PTOLEMY. 

Looking ahead, there is wide scope for further investigations of the scenarios proposed here and the general DM stability mechanism. In particular, it would be very interesting to explore the detailed predictions for precision cosmological and astrophysical observables through, e.g., the study of cosmological perturbations and/or structure formation, which will help to more accurately delineate the constraints and future prospects for the various probes discussed in this paper. Likewise, for the scenario in which DM is stabilized by neutrinos, it would be valuable to carry out more detailed phenomenological investigations of the properties of the present-day C$\nu$B and its detection prospects, including the exploration of more general  flavor patterns of the $\phi$-neutrino couplings. Finally, it would be worthwhile to explore other models realizing the mechanism of DM stability by Pauli blocking and their phenomenology.

\section*{Acknowledgements}
W.Y. thanks the Department of Physics and Astronomy of the University of Pittsburgh for its hospitality when this work was initiated in 2019. 
B.B. is supported in part by the U.S. Department of Energy under grant No. DE-SC0007914 and in part by PITT PACC. W.Y. was supported by JSPS KAKENHI Grant Nos. 16H06490, 20H05851, 21K20364, 22K14029, 22H01215, 23K22486 and by NRF Strategic Research Program NRF-2017R1E1A1A01072736.

\clearpage

\appendix

\section{Generic formulas for decay and scattering}

\label{app:1}
In the main text, we have focused on real scalar as the DM interacting with a Majorana fermion. 
In this appendix, we provide the relevant formulas for decay and scattering reactions for both scalar and pseudoscalar couplings. 
To this end, we consider the Lagrangian interaction
\begin{equation}
\laq{Lgene}
{\cal L} \supset - \frac{1}{2} \phi  \,\overline \psi ( y_S + y_P i \gamma^5) \psi,
\end{equation}
where $y_S$ and $y_P$ are both real. Here $\psi$ describes a four-component Majorana fermion. 

\subsection{Decay $\phi \rightarrow \psi \psi $}
The partial decay width for $\phi \rightarrow \psi \psi $ is 
\begin{equation}
\label{eq:phi-decay-width}
\Gamma_{\phi\to \psi\psi} = \frac{1}{2!} \frac{m_\phi}{8 \pi} \left( y_S^2 \beta_\psi^3 + y_P^2 \beta_\psi \right), ~~~~~~~~~~ \beta_\psi = \left(1-\frac{4 m_\psi^2}{m_\phi^2}\right)^{1/2}.
\end{equation}
We note that for the scalar coupling, the decay width exhibits a $P$-wave suppression, while for the pseudoscalar coupling, the decay proceeds in the $S$-wave.

\subsection{Annihilation $\phi \phi \rightarrow \psi \psi$}
Here we provide some details for the annihilation process $\phi \phi \rightarrow \psi \psi$ for the case of scalar and pseudoscalar couplings. 

\subsubsection{Scalar coupling}

Consider the Lagrangian interaction{, i.e. $y_S=y, y_P=0$ in \Eq{Lgene},}
\begin{equation}
{\cal L} \supset - \frac{y}{2} \phi  \overline \psi \psi.
\end{equation}
The amplitude for the annihilation process is 
\begin{equation}
{\cal M} = - y^2 \, \overline u(p_3) \left[ \frac{ \not{\! p}_3 - \not{\! p}_1 + m_\psi   }{t-m_\psi^2}+\frac{ \not{\! p}_1 - \not{\! p}_4 + m_\psi   }{u-m_\psi^2} \right] v(p_4),
\end{equation}
where we have introduced the Mandelstam variables with the standard definitions, $s = (p_1+p_2)^2$, $t = (p_1-p_3)^2$, $u = (p_1-p_4)^2$.
{The  spin summed squared amplitude} is 
\begin{equation}
\label{eq:scalar-ann-M2}
|\overline {\cal M}|^2  = y^4 \times 2 \left[ \frac{F_{tt}}{(t-m_\psi^2)^2} + \frac{F_{uu}}{(u-m_\psi^2)^2} + \frac{F_{tu}+F_{ut} }{(t-m_\psi^2)(u-m_\psi^2)}   \right],
\end{equation}
where 
\begin{align}
F_{tt} & = t \, u- m_\psi^2 \, (9\, t + u) - 7\, m_\psi^4 + 8\, m_\psi^2 \, m_\phi^2 - m_\phi^4, \nonumber \\
F_{uu} & = t \, u- m_\psi^2 \, (9\, u + t) - 7\, m_\psi^4 + 8\, m_\psi^2 \, m_\phi^2 - m_\phi^4, \\
F_{tu}= F_{ut} & = -t \, u -3\, m_\psi^2 \, ( t + u) - 9\, m_\psi^4 + 8\, m_\psi^2 \, m_\phi^2 + m_\phi^4. \nonumber
\end{align}
The differential cross-section can be written as 
\begin{align}
d \sigma 
= \frac{1}{2!} \frac{1}{16 \pi \beta_\phi^2 s^2} |\overline {\cal M}|^2 dt,
\end{align}
where 
$\beta_i = (1-4m_i^2/s)^{1/2}$.

The formula for the integrated cross-section in full generality is complicated and we will not write it here. However, it is interesting to consider the nonrelativistic limit and compute the quantity $\sigma v_{\rm rel}$,
where $v_{\rm rel} = |v_1 - v_2| = 2 \beta_\phi$ is the relative velocity of the $\phi$ particles in the CM frame. In the non-relativistic limit we obtain
\begin{equation}
\sigma v_{\rm rel} \simeq  \frac{2 y^4 m_\psi^2}{\pi m_\phi^4} \left(1- \frac{m_\psi^2}{m_\phi^2}\right)^{3/2}.
\end{equation}
We see that the annihilation proceeds through the $S$-wave, but exhibits a chirality suppression. The implications of the two cases are explored further in Section \Sec{DMODS}.

\subsubsection{Pseudocalar coupling}

Consider the Lagrangian interaction,  i.e. $y_S=0, y_P= y$ in \Eq{Lgene},
\begin{equation}
{\cal L} \supset - i \frac{y}{2} \phi  \overline \psi \gamma^5 \psi .
\end{equation}
The amplitude of the annihilation process is 
\begin{equation}
{\cal M} =  y^2 \, \overline u(p_3) \, \gamma^5 \, \left[ \frac{ \not{\! p}_3 - \not{\! p}_1 + m_\psi   }{t-m_\psi^2}+\frac{ \not{\! p}_1 - \not{\! p}_4 + m_\psi   }{u-m_\psi^2} \right] \, \gamma^5 \, v(p_4) ,
\end{equation}
We see the appearance of the $\gamma^5$ now. The spin-averaged squared amplitude is 
\begin{equation}
|\overline {\cal M}|^2  = y^4 \times 2 \left[ (t-m_\psi^2)(u-m_\psi^2)  -m_\phi^4 \right]  \left[ \frac{1}{t-m_\psi^2} - \frac{1}{u-m_\psi^2}   \right]^2 .
\end{equation}
The cross-section is obtained from the general formula listed in the previous subsection. In the non-relativistic limit we obtain
\begin{equation}
\sigma v_{\rm rel} \simeq  v_{\rm rel}^4 \, \frac{ y^4}{60 \pi m_\phi^2} \left( 1 +  \frac{3m_\psi^2}{2 m_\phi^2}  \right)
 \left(1- \frac{m_\psi^2}{m_\phi^2}\right)^{3/2} .
\end{equation}
The annihilation proceeds through the $D$ wave but does not exhibit a chirality suppression. While we have note studied the pseudoscalar coupling in detail in this work, we expect this scenario to give qualitatively similar features to the scalar coupling in the chirality suppressed cased studied in \Sec{cs}.

\section{Limits on fermion energy in collision integrals}
\label{app:2}

\subsection{Decay $\phi(p_\phi) \rightarrow \psi(p_1) \psi(p_2)$}

In this collision term (\ref{eq:C-phi-decay-th})
 we encounter an integral over the fermion energy $E_1$. 
 The limits are determined from the energy conserving delta function for $\cos\theta = \pm 1$, leading to the condition
\begin{equation}
0 = E_\phi - E_1 - \sqrt{m_\psi^2 + p_\phi^2+ p_1^2 \pm 2 p_\phi p_1 } .
\end{equation}
Solving this equation, we obtain
\begin{align}
\label{eq:int-limits-decay}
p_1^\pm & = \frac{1}{2} (E_\phi \beta_\psi \pm p_\phi), \\
E_1^\pm & = \frac{1}{2} (E_\phi \pm \beta_\psi p_\phi),
\end{align}
where $\beta_\psi$ is given in Eq.~(\ref{eq:phi-decay-width}).

\subsection{Annihilation $\phi(p_\phi) \phi(p_2) \rightarrow \psi(p_3) \psi(p_4)$}
In this collision term (\ref{eq:C-phi-ann-cth})
 we encounter an integral over the fermion energy $E_3$. As above, the limits of integration follow from the energy conserving delta function when $\cos\theta = \pm 1$, leading to the equation
\begin{equation}
0 = m_\phi + E_2  - E_3 - \sqrt{m_\psi^2 + p_2^2+ p_3^2 \pm 2 p_2 p_3 } .
\end{equation}
Solving this equation, we obtain
\begin{align}
\label{eq:int-limits-ann}
p_3^\pm & = \frac{1}{2} [ (E_2+m_\phi) \kappa_\psi \pm p_2], \\
E_3^\pm & = \frac{1}{2} (E_2+m_\phi \pm \kappa_\psi \, p_2),
\end{align}
where in this case 
\begin{equation}
\kappa_\psi = \left(1-\frac{2m_\psi^2}{m_\phi (E_2 + m_\phi)}\right)^{1/2}.
\end{equation}

\section{Renormalizable model for DM stabilized by C$\nu$B }
\label{sec:ap1}
In this appendix we present a renormalizable UV model realizing the scenario discussed in \Sec{DMONS}. Along with the scalar $\phi$ which serves as DM, the theory contains a Dirac fermion mediator $N$. The Lagrangian shares some features with the Dirac-neutrino portal model~\cite{Bertoni:2014mva,Batell:2017cmf, Yin:2018yjn},
\beq
{\cal L}_{\rm int}=
\frac{y_1}{2} \f \bar{{N}}^c_L  N_L+\frac{y_2}{2} \f \bar{{N}}^c_R  N_R + M_N \bar{N}_R N_L + y_R \bar{N}_R H   L
+ {\rm h.c.}~
\eeq
The model enjoys an approximate
lepton $Z_4$ parity where $(\f, N_R, N_L, L, H)$ are charged as $(2, 1, 1, 1,0)$. 
 By integrating out the Dirac right-handed neutrinos
one obtains the effective Lagrangian,
\beq
{\cal L}_{\rm eff} \supset \frac{y}{2} \f\bar{\nu}_L^c\nu_L+{\rm h.c.} 
\eeq
In  small $y_2$ regime, we obtain $y = y_1 y_R^2 v^2/2 M_N^2$
at energies much below the electroweak scale $v$.
Notice that the neutrino Majorana mass term $\tfrac{1}{2}m_\nu \overline\nu^c_L \nu_L$ is forbidden by the $Z_4$ symmetry. 
To generate neutrino masses, one can introduce a small lepton parity violating term. 

Setting aside possible model-dependent UV contributions, the radiative corrections to the $\f$ mass from the new degrees of freedom in this model are given by
\beq
\d m_\f^2\sim \frac{y_{1}^2 M_N^2}{16\pi^2}.
\eeq
For example, taking $y_{R} v \sim M_N$ such that $y \sim y_1$,
we obtain
\beq
\d m_\f^2\sim \(10^{-4}\EV\)^2 \(\frac{y}{10^{-14}}\)^2 \(\frac{|M_N|}{100\GEV}\)^2. 
\eeq

In the early Universe, especially in the symmetric phase, the thermalization rate of $\f$ particle is estimated as  
\beq
\G_{\rm th}\sim \frac{y_{1}^2 y_{R}^2}{4\pi}T.
\eeq
This is much smaller than the expansion rate of the Universe for
$T\sim 100\GEV\AND y_{1}\sim y$ in the parameter region in Fig.\ref{fig:DMONS}. 
On the other hand, the induced  $\phi$ thermal mass may be larger than the bare mass depending on the parameter region. In this case, $\phi$ condensate behaves as radiation at the onset of oscillation, rather than matter because the effective mass scales with $a^{-1}$. However, our overall conclusions do not change since we can require that at later times $\phi$ accounts for the DM once the thermal mass becomes smaller than the bare mass due to the redshift.

\bibliography{references}
\end{document}